\documentclass[journal]{IEEEtran}
\usepackage{ulem}
\usepackage{algorithm}
\usepackage{algorithmic}
\usepackage{helvet}
\usepackage{courier}
\usepackage{subfigure}
\usepackage{graphicx}
\usepackage{bm}
\usepackage{cite}

\usepackage{amsmath}
\usepackage{epstopdf}
\usepackage{multirow}
\usepackage{color}
\usepackage{booktabs} 
\usepackage{enumerate}
\usepackage{paralist, blindtext}

\ifCLASSINFOpdf
\else
\fi
\usepackage[justification=centering]{caption}

\begin{document}
%
\title{A Structural Representation Learning for Multi-relational Networks}
%
%
%
\author{Xin~Li,
        Huiting~Hong,
        Lin~Liu
        and~William K.~Cheung
}

\maketitle

\begin{abstract}
Most of the existing multi-relational network embedding methods, e.g., TransE, are formulated to preserve pair-wise connectivity structures in the networks. With the observations that significant triangular connectivity structures and parallelogram connectivity structures found in many real multi-relational networks are often ignored and that a hard-constraint commonly adopted by most of the network embedding methods is inaccurate by design, we propose a novel representation learning model for multi-relational networks which can alleviate both fundamental limitations. Scalable learning algorithms are derived using the stochastic gradient descent algorithm and negative sampling. Extensive experiments on real multi-relational network datasets of WordNet and Freebase demonstrate the efficacy of the proposed model when compared with the state-of-the-art embedding methods.
\end{abstract}


\IEEEpeerreviewmaketitle

\section{Introduction}
\IEEEPARstart{R}{epresentation} learning has become an important research track in the area of machine learning, with the aim of providing more informative numerical representations of the observed data
for applications like image classification, speech recognition and text mining, etc. More specifically, network embedding, which is to learn the distributed representations of information networks, has attracted much attention due to the promising empirical results obtained.
In the literature, a number of network embedding methods have been proposed, including LINE \cite{DBLP:conf/www/TangQWZYM15}, IONE \cite{DBLP:conf/ijcai/LiuCLL16}, SDNE \cite{DBLP:conf/kdd/WangC016}, and DeepWalk \cite{DBLP:conf/kdd/PerozziAS14}.
These methods learn only the representations of the nodes in a network, and the edges are assumed to be single-relational, that is, they are of the same type. For instance, edges represent only ``friendship'' in a social network , and only ``collaboration'' in the DBLP collaboration network.

A multi-relational network is represented by a directed graph with the edges of various relation types typically indicated by associating each edge from a source node to a target node with a discrete label, denotes as \textbf{\emph{(source, label, target)}} or \textbf{\emph{(h, r, t)}}. Such multi-relational networks, e.g., Google Knowledge Graph, semantic networks and multi-relational social networks, have become important resources to support more advanced information retrieval, question-answering systems, etc.
To learn the embedding of such a network, it is common for both the node and edge representations to be learned at the same time.



\begin{figure}[t]
	\centering
	\subfigure[]{\includegraphics[width=1.5in]{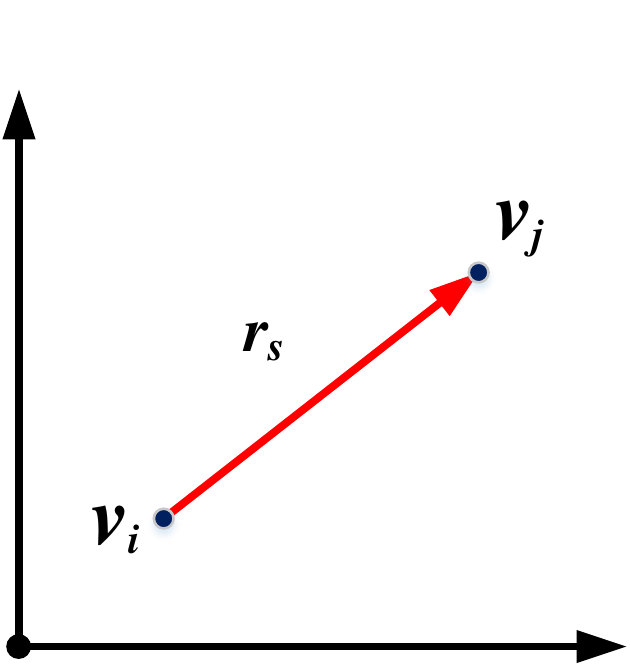}
     \label{fig:triangle1} }
	\subfigure[]{\includegraphics[width=1.5in]{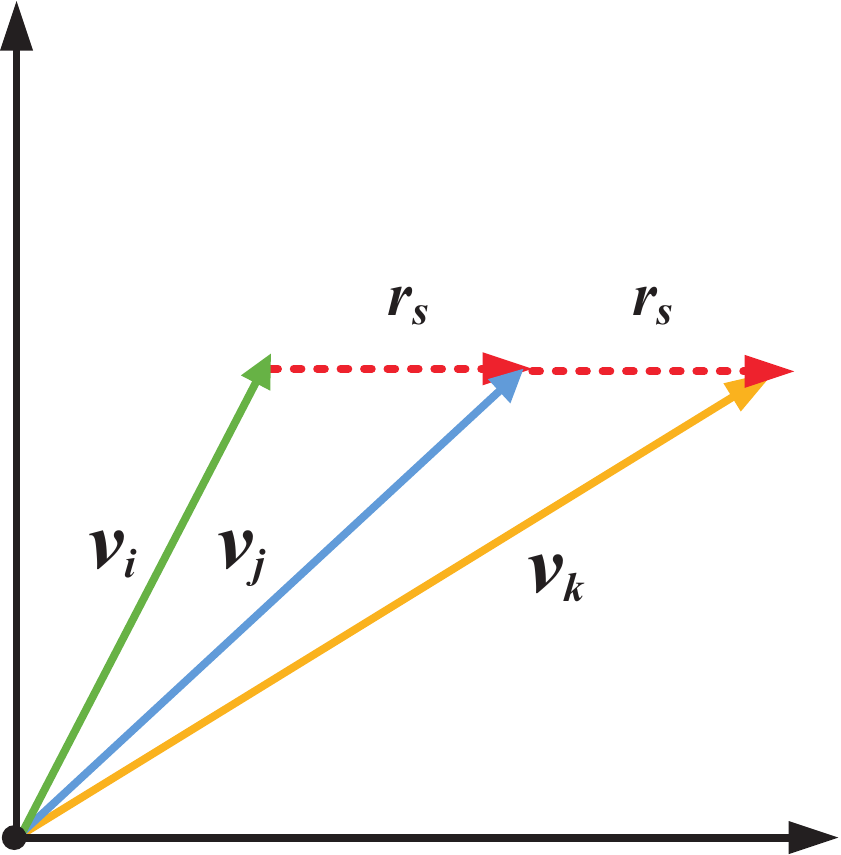}
		\label{fig:triangle2} }
	\caption{Trans-family vs. Triangular structures}\label{fig:triangle}
\end{figure}

Following the success of TransE \cite{DBLP:conf/nips/BordesUGWY13}, a series of translation-based methods have been proposed for knowledge graph (KG) embedding to project the nodes (also called entities) and the edges (also called relations) of the KG onto a continuous vector space, e.g., TransH \cite{DBLP:conf/aaai/WangZFC14}, TransR \cite{DBLP:conf/aaai/LinLSLZ15}, pTransE\cite{DBLP:conf/emnlp/WangZFC14} and TransG \cite{DBLP:journals/corr/0005HHZ15} (referred to as ``trans-family'' hereafter), so that the local structural relationship of the nodes and edges can be retained in their corresponding embeddings. These approaches differ from each other in the way of (1) whether the entities and relations are projected onto the same subspace (e.g., TransH and TransR project a KG onto different subspaces to reflect the relations' semantics); (2) how the embedding objective function is defined (e.g., TransE minimizes the so-called energy function of $f_r\left( h,t \right)=||h+r-t||$ , while pTransE maximizes the conditional probability of $\left( h,r,t \right)$ with the constraint $h+r=t$).

\begin{figure*}[t]
	\centering
	\subfigure[]{\label{tran1}\includegraphics[width=0.25\paperwidth]{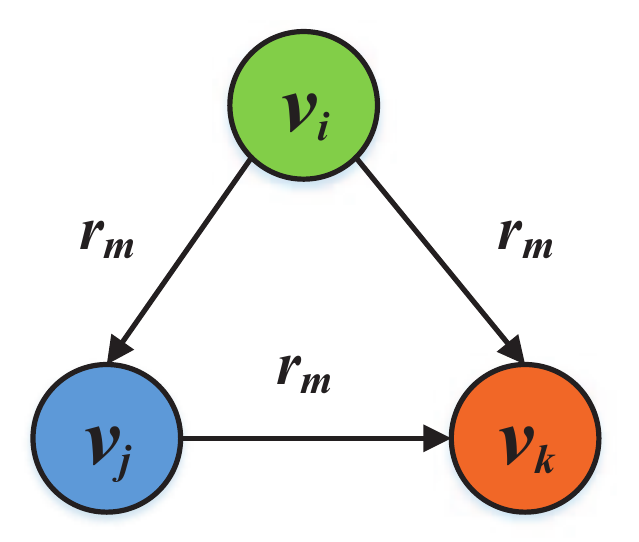}}
	\subfigure[]{\label{tran2}\includegraphics[width=0.25\paperwidth]{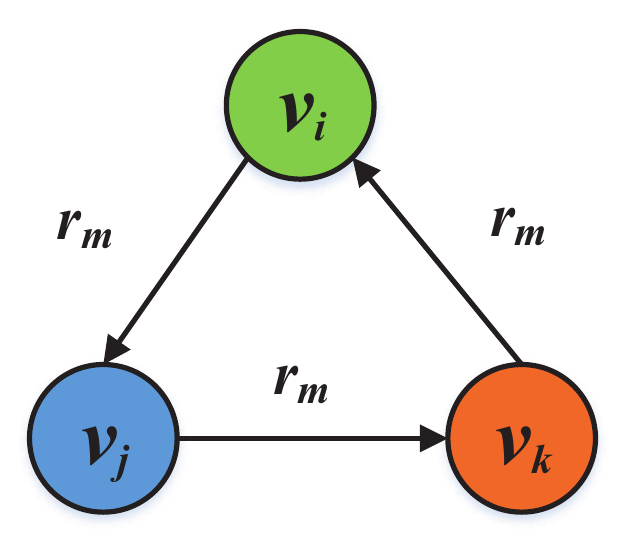}}
	\\
	\subfigure[]{\label{tran3}\includegraphics[width=0.25\paperwidth]{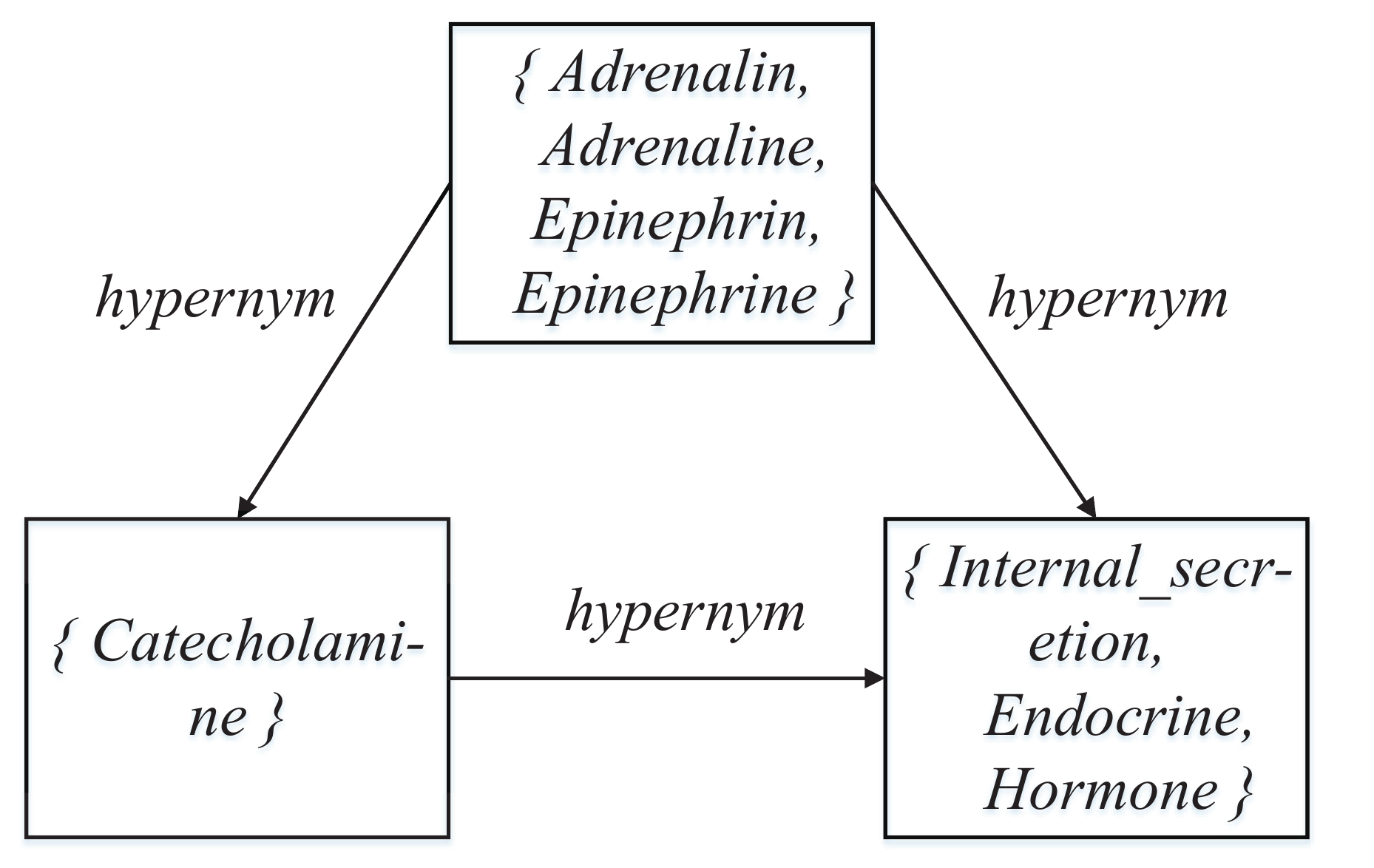}}
	\subfigure[]{\label{tran4}\includegraphics[width=0.25\paperwidth]{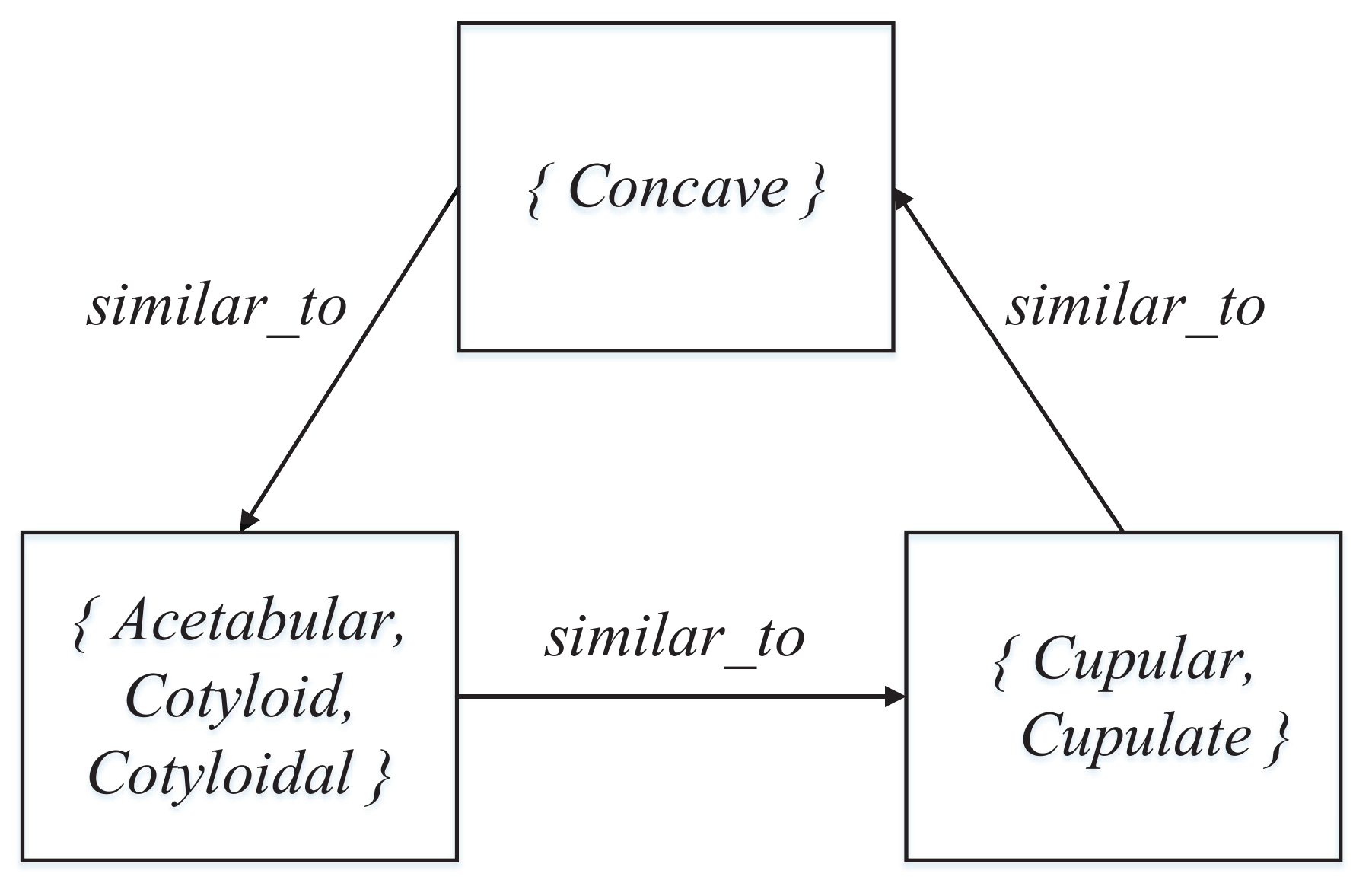}}
	\subfigure[]{\label{tran5}\includegraphics[width=0.25\paperwidth]{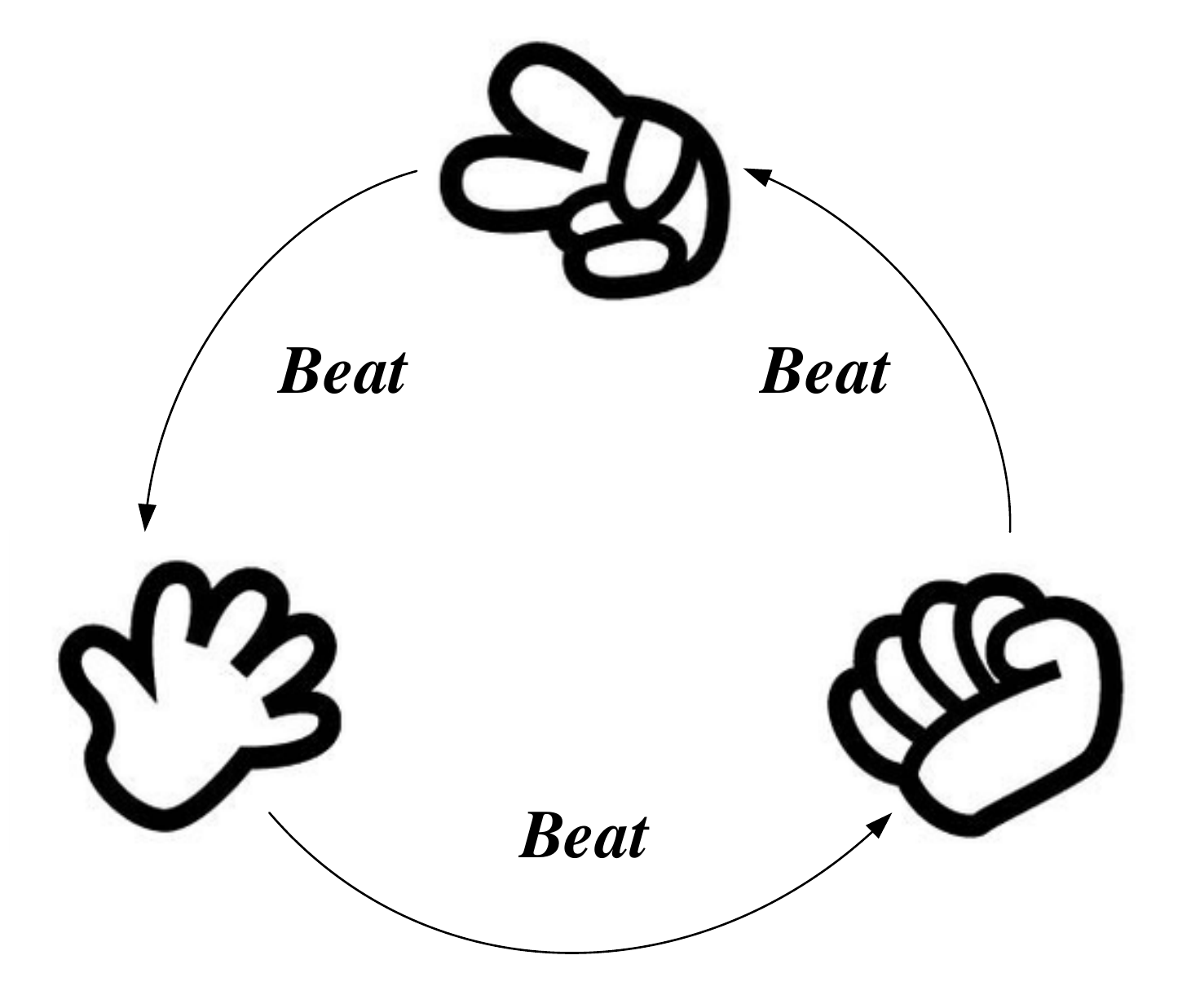}}
	\caption{Triangular structure examples}
    \label{tranmode}
\end{figure*}


In this paper, we focus on the representation learning of the multi-relational networks, and propose our approach based on the following two observations:\\


\textbf{Observation 1:} Methods in the trans-family are all constrained by $h+r=t$ which cannot capture the structures shown in Fig.\ref{tranmode}. For the directed graph with three nodes connecting to each other via a specific edge, there are two non-isomorphic modes. In the trans-family, the scoring function ${{f}_{r}}\left( h,t \right)=||h+r-t||$ is used to ensure the plausibility of triple  $\left( h,r,t \right)$. Accordingly, the closeness of similar nodes can be guaranteed in the low-dimensional Euclidean space. However, Euclidean geometry breaks when encountering triangular structures. For example, TransE requires the forms of $v_i+r_m\approx v_j$, $v_j+r_m\approx v_k$ and $v_i+r_m\approx v_k$ to hold at the same time. However, as illustrated in Fig.\ref{fig:triangle1}, for the former two equations to hold, we have $v_i+2r_m\approx v_k$. The forcible updating rule in trans-family will compromise the accuracy. In this paper, this structure is referred to as the triangular structure which often appears in many multi-relational networks. For example, Fig.\ref{tran3} illustrates a fact in WordNet which accords with the mode in Fig.\ref{tran1}, where \{internal\_secretion, endocrine, hormone\} is the hypernym of \{adrenalin, adrenaline, epinephrin, epinephrine\} and \{catecholamine\}, \{catecholamine\} is the hypernym of \{adrenalin, adrenaline, epinephrin, epinephrine\}, and the relation edge is labeled ``hypernym''.  Note that WordNet is organized by the concept of synonym sets (so-called synsets), where each node represents a set of words that are roughly synonymous in a given context. Fig.\ref{tran4} illustrates another fact in WordNet which accords with the mode in Fig.\ref{tran2}, where the relation is ``similar to''. When the relation $r_m$ comes to ``similar to'', someone can argue that $r_m$ can be set as a zero vector so that the constraints of $h+r=t$ hold within the triangular structures, leading the representations of nodes are similar to each other as the constraints have been transformed into the form of $h=t$. However, a similar argument cannot be made for other type of relations. For example, Fig.\ref{tran5} shows a well-known hand game which accords with the mode in Fig.\ref{tran2}. In the game, rocks beat/defeat scissors, scissors beat/defeat papers and papers beat/defeat rocks. Obviously, it is not appropriate to set the relation of beat as zero vectors while leading rocks, scissors and papers have the same low-dimensional representations.

\begin{figure}[t]
	\subfigure[]{\label{fig:pingxing1}\includegraphics[width=3.3in]{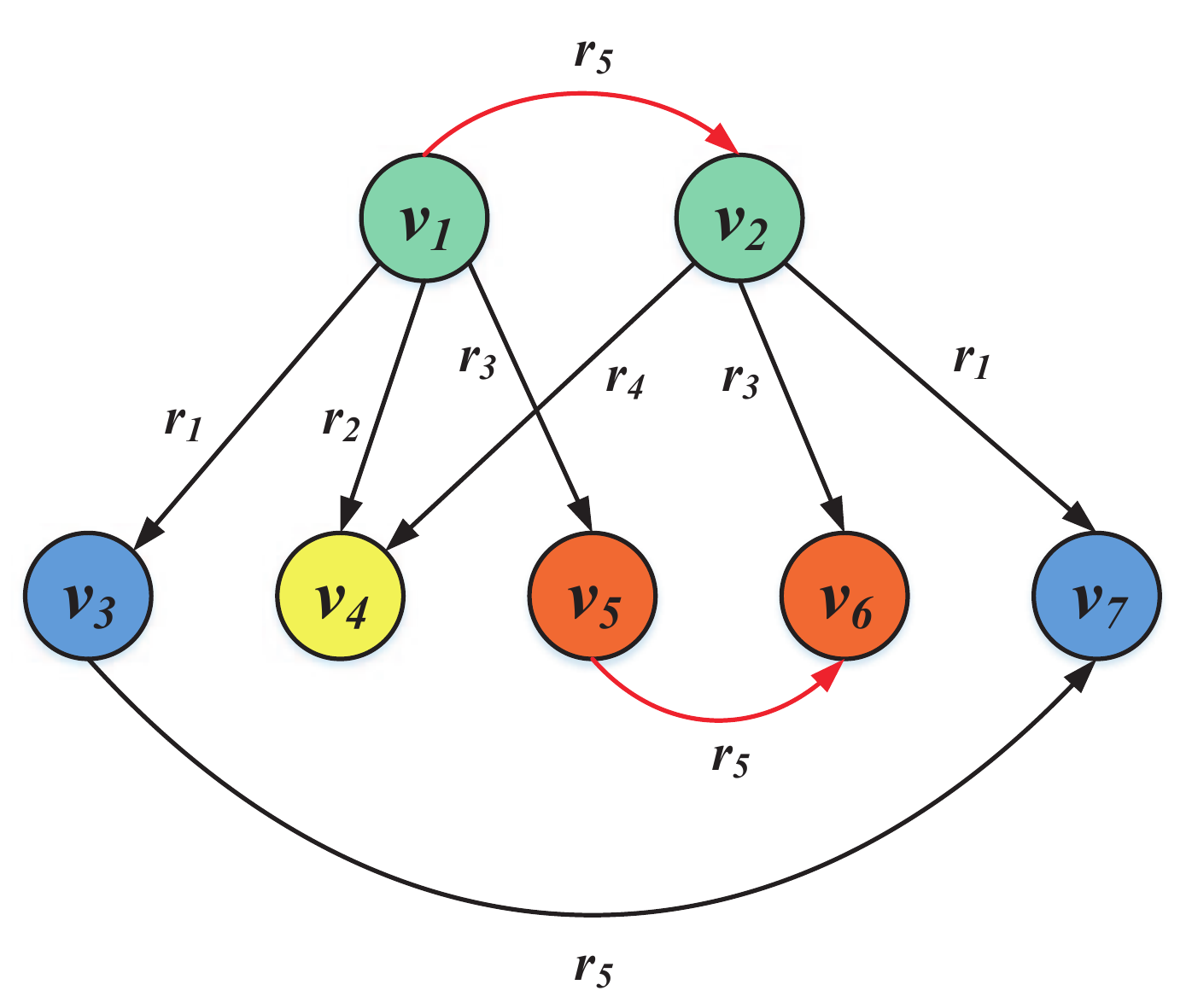}}
	\subfigure[]{\label{fig:pingxing2}\includegraphics[width=3.3in]{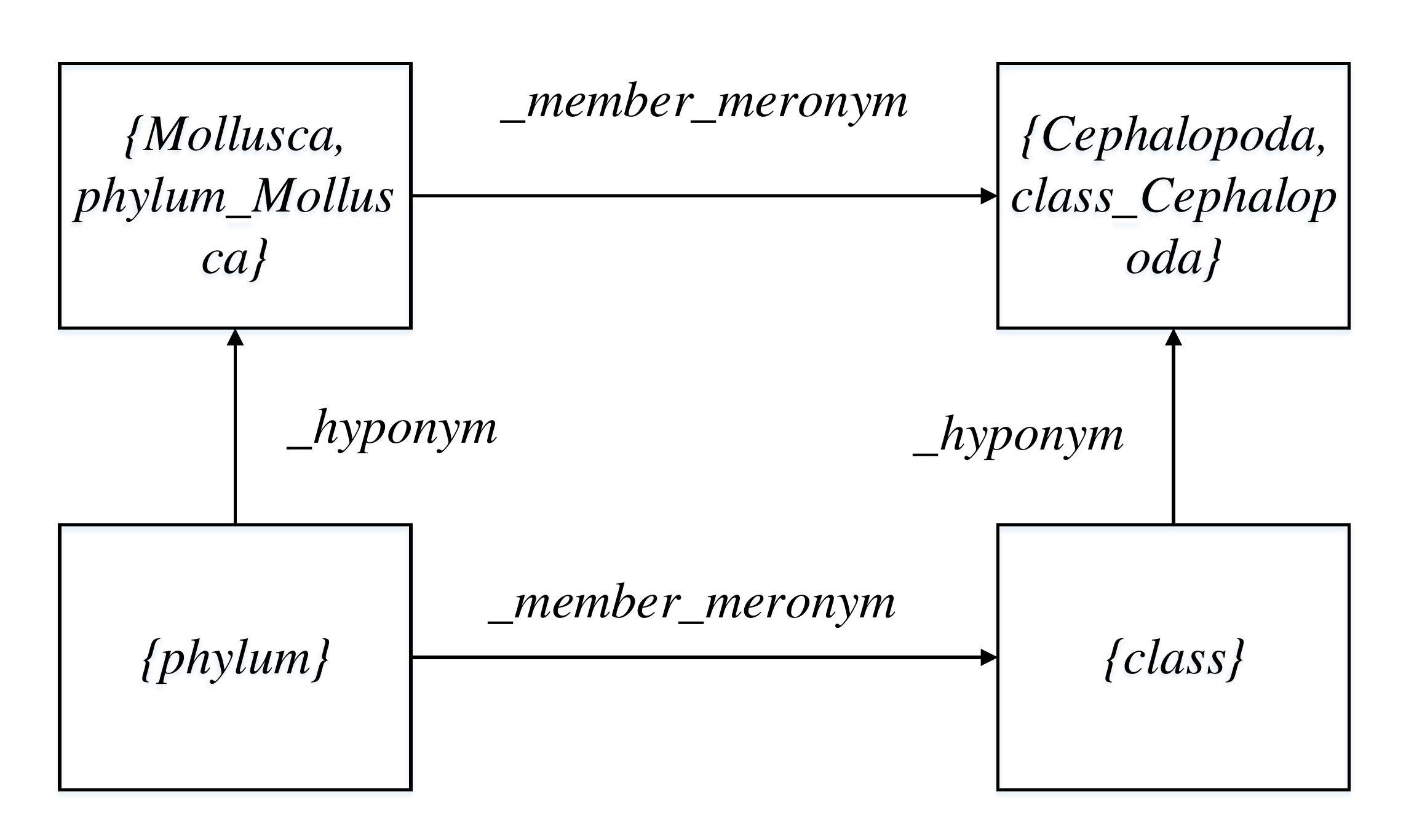}}
	\caption{Parallelogram structure examples.}
  \label{parallel}
\end{figure}

\textbf{Observation 2:} Network embedding methods like LINE \cite{DBLP:conf/www/TangQWZYM15} have been proposed to capture network structures by exploring the first-order and second-order proximities. The former corresponds to the edge strength between two connected nodes, while the latter corresponds to the overlapping neighbors of the two nodes. Note that embedding methods like LINE are deliberately designed for single-relational networks in which these two properties are commonly seen. However, in multi-relational networks, the strengths of the edges do not vary as much as in single-relational networks\footnote{For example, the number of mentions(@) of a user by another user could be considered as the strength of these two users (nodes) in single relational social networks.}. For KGs like WordNet, most of nodes are linked with each other by an edge of a specific relation type only once. Besides, it is difficult to define the scale of the strength when the relations have different semantic meanings. In addition, the second-order proximity focuses on how many neighbors of two nodes are exactly the same, whereas in our framework we propose to relax such proximity definition by considering the proximity among the neighbors via {\it parallelogram structures}. We have found that parallelogram structures exist more often in multi-relational networks. Fig.\ref{fig:pingxing1} illustrates the examples of parallelogram structures, where $\{v_1,v_2,v_5,v_6\}$ and $\{v_1,v_2,v_3,v_7\}$ are the two instances of the parallelogram structure with the parallel sides of the same relation type.
Intensively, the two nodes $v_1$ and $v_2$ are linked to $v_3$ and $v_7$ via the same relation $r_1$ respectively. When $v_3$ and $v_7$ are linked by a relation $r_5$, it is highly likely $v_1$ and $v_2$ can be linked together via the same relation of their neighbors, that is $r_5$. Intuitively, given any three sides of the parallelogram, we could infer the relation of the fourth one. Fig.\ref{fig:pingxing2} is an instance in WordNet which accords with a parallelogram mode, where \{Cephalopoda, class\_Cephalopoda\} and \{Mollusca, phylum\_Mollusca\} are hyponyms of \{class\} and \{phylum\} respectively. If you also know that \{Cephalopoda, class\_Cephalopoda\} is a member of \{Mollusca, phylum\_Mollusca\}, it is undoubtedly logical that \{class\} is a member of \{phylum\}.

In this paper, we propose a multi-relational network embedding method. The objective function is designed to consider deliberately the triangular and parallelogram structures to define node proximity, and thus to infer the representations. In order to improve the efficiency, we adopt the stochastic gradient descent algorithm and negative sampling to optimize the objective function to reduce the training cost. We conduct extensive experiments over the tasks of triplet classification and  link prediction on the real datasets like WordNet and Freebase. Experimental results demonstrate the effectiveness of our model over several state-of-the-art methods.

%
%
%

\section{Related Work}
There are two lines of research related to our work, namely {\it network embedding} and {\it knowledge graph embedding}.
\subsection{Network Embedding}

One of the recent attempts to address network embedding is graph factorization (GF) \cite{DBLP:conf/www/AhmedSNJS13} which utilizes matrix factorization over undirected graph's affinity matrix to infer the low-dimensional embedding. Only first-order proximity is preserved and nodes with close interaction are represented closely in the projected vector space. LINE \cite{DBLP:conf/www/TangQWZYM15} is another recently proposed method to handle large-scale network embedding for both directed and undirected graphs, where both first-order and second-order proximity measures have been considered. DeepWalk \cite{DBLP:conf/kdd/PerozziAS14}
utilizes the distribution of node degree to model network community structure via random walk and skip-gram together to infer the network embedding. However, the studies show that DeepWalk tends to preserve the second-order proximity only. HARP\cite{DBLP:journals/corr/ChenPHS17} is proposed as a general meta-strategy to improve the graph representation learning methods, such as LINE and Deepwalk, by collapsing edges to gain the coarse graphs for higher-order graph structural information. However, the way of edge collapsing makes it difficult for adapting HARP on multi-relational networks.
SDNE \cite{DBLP:conf/kdd/WangC016} offers a semi-supervised deep learning framework to address the problem of learning representations of networks, in which the first-order proximity and the second-order proximity are jointly preserved. The existing network embedding methods mainly focus on networks with pairwise relationships. While DHNE\cite{DBLP:journals/corr/abs-1711-10146}  switches the attention to tuple-wise relationships, which is defined as hyperedges in the hyper-network. Practically, DHNE combines the multilayer perceptron and the autoencoder to model the tuplewise similarity function and preserve both local and global proximities in the formed hyper-network embedding space.

Recently, Generative Adversarial Nets(GAN) by designing a game-theoretical minimax game have received a great deal of attention. Inspired by GAN, AIDW\cite{DBLP:journals/corr/abs-1711-07838} introduces GAN on the basis of Deepwalk to guarantee embedding learned satisfy prior distribution for learning robust graph representations.
GraphGAN\cite{DBLP:journals/corr/abs-1711-08267} is another recently proposed approach, where the discriminator tries to distinguish well-connected vertex pairs from ill-connected ones and graph softmax is proposed as the implementation of the generator to solve the inherent limitations of the traditional softmax. However these adversarial approaches are notorious for their unstable training process.   

Furthermore, the above methods usually study networks with a single type of proximity between nodes, which defines a single view of a network. However, in practice there usually exists multiple types of proximities between nodes, yielding networks with multiple views.
MVE\cite{DBLP:conf/cikm/QuTSR0017} regards the multi-type network as multiple single-relational(single-view) networks and studies the node representations for networks with multiple views on the same semantic vector space. The node representations across different views can be obtained by summing up the weighted embeddings of node on all single-view networks.
PTE\cite{DBLP:conf/kdd/TangQM15} is a semi-supervised method to handle the embedding of the multi-type networks, where the nodes are of different types. PTE divides the network into multiple sub-networks according to the type of nodes to learn each sub-networks embedding by using LINE. In particular, the same nodes in different sub-networks share the same embedding.

In summary, most existing network embedding  approaches learn the representations of nodes in single-relation networks, or transform the representation tasks of multi-relational networks into single-relational network embedding tasks. The semantics of multiple relations are also not addressed in multi-type networks. Besides, as explained in Section I, the first-order and second-order proximity adopted in most existing work may not be the representative local structures in multi-relational networks.

\subsection{Knowledge Graph Embedding}
Recent advance of relational learning for knowledge graph embedding has attracted much attention from industry and academia. Among them, TransE \cite{DBLP:conf/nips/BordesUGWY13} is the most well-known pioneer work which embeds both nodes and edges of different relation types onto a low-dimensional vector space. The basic idea is to represent the edge (relation) of two nodes (entities) as a translation operation in the embedding space. Given the triplet $\left( h,r,t \right)$, we expect the representation vector of the node $t$ to be as close as possible to the representation vector of the node $h$ plus the relation $r$. The objective function is $||h+r-t||$. TransE is an efficient algorithm for the embedding. However, it does not do well in dealing with some mapping properties of relations, such as reflexive, one-to-many, many-to-one, and many-to-many. 

To alleviate the limitations, Wang {\it et al.} proposed TransH \cite{DBLP:conf/aaai/WangZFC14} to project the nodes in a relation-specific subspace (a hyperplane ${{w}_{r}}$) to obtain ${h}'$ and ${t}'$ respectively for each triplet $\left( h,r,t \right)$. The translation is performed in the relation subspace and constrained by the function of ${h}'+r={t}'$.
Lin {\it et al.} extended the idea of TransH and proposed TransR \cite{DBLP:conf/aaai/LinLSLZ15} to project the entities and relations onto different vector spaces respectively to further increase the degrees of freedom for the representations. To adapt various mapping properties, TransM\cite{DBLP:conf/paclic/FanZCZ14} was proposed to leverage on the structures of the knowledge graph by pre-calculating the distinct weight for each training triplet with respect to different relational mapping property. TransH and TransM only consider ``one hop'' information about directed linked entities while missing more global information.

While, in \cite{DBLP:conf/emnlp/LinLLSRL15}, the authors argued that multiple-step relation paths also contain rich inference patterns between entities, and proposed a path-based representation learning model by considering relation paths as translations between entities. In addition to path information, neighbor context and edge context are introduced by GAKE\cite{DBLP:conf/coling/FengHYZ16} to reflect the property of knowledge graph from different perspectives.

Wang {\it et al.}\cite{DBLP:conf/emnlp/WangZFC14} proposed a probabilistic TransE to encode the knowledge graph by maximizing the conditional probability of $\left( h,r,t \right)$, in which the conventional scoring function of $||h+r-t||$ is still being utilized. 
TorusE\cite{DBLP:journals/corr/abs-1711-05435} introduces the torus, which is one of the Compact Lie Groups, to replace the regularization term of the conventional TransE to obtain a more robust link prediction.



These translation-based approaches inherit the efficiency from TransE but also the underlying flaws when using the scoring function in one way or another. As illustrated in Section I, the use of the constraint of $h+r=t$ cannot handle the triangular structures of multi-relational networks. In this paper, we propose a novel multi-relational network embedding approach to overcome the flaws of the trans-family where the observed local structures are incorporated into the objective function to infer a more robust network representation.
\begin{figure}[t]
	\centering\includegraphics[width=3in]{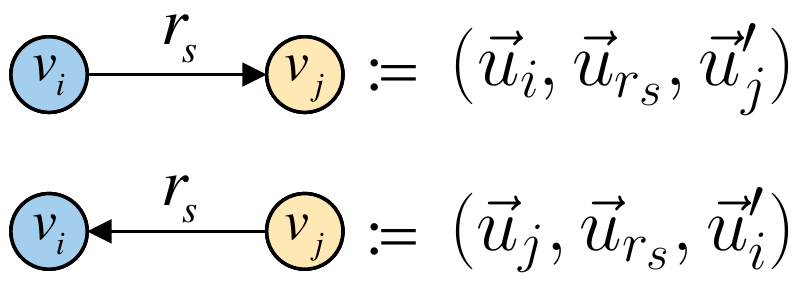}
	\caption{Vector representation for directed networks}\label{fig:directedEdge}
\end{figure}
\begin{figure*}[t]
	\centering
	\subfigure[]{\includegraphics[width=2.1in]{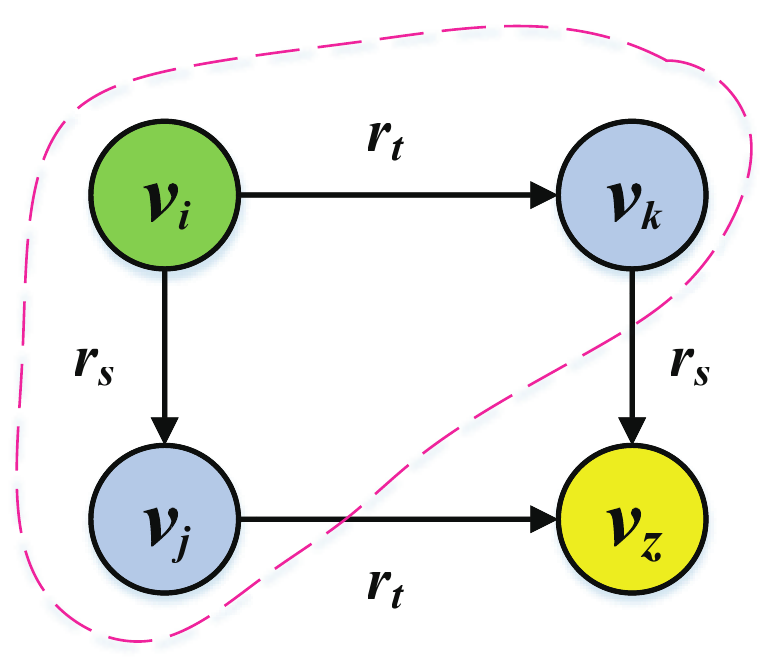}
     \label{fig:case1}}
	\subfigure[]{\includegraphics[width=2.1in]{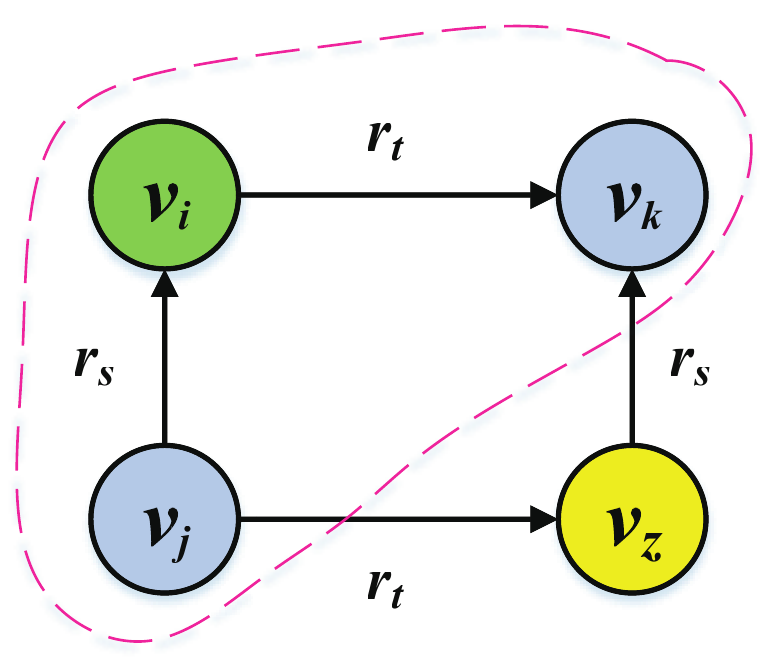}
		\label{fig:case2}}
	\subfigure[]{\includegraphics[width=2.1in]{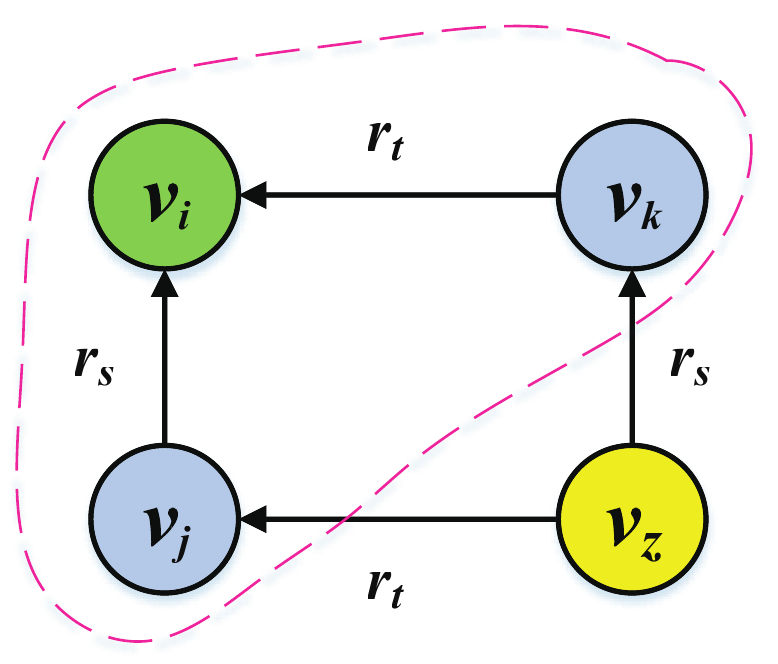}
		\label{fig:case3} }
	\caption{Local connectivity structures of parallelogram}\label{fig:cases}
\end{figure*}

\section{Model Framework}

Let $G=( V, E, R)$ be the graph representation of a directed multi-relational network where $V=\{ {{v}_{1}},{{v}_{2}},\ldots ,{{v}_{|V|}} \}$ corresponds to the set of nodes, $R=\{ {{r}_{1}},{{r}_{2}},\ldots ,{{r}_{|R|}} \}$ corresponds to the set of relation labels, and $E$ corresponds to the set of typed edges. Each typed edge in $E$ is denoted as a triplet $( {{v}_{i}},{{r}_{s}},{{v}_{j}} )$ with ${{v}_{i}}$ being the source node, ${{r}_{s}}$ being the associated relation label, and ${{v}_{j}}$ being the target node.



\subsection{Model Description}
\label{sec:model_description}
We propose a novel probabilistic embedding model for representing multi-relational networks. Similar to most of existing representation learning methods, we represent each node ${{v}_{i}}\in V$ as a d-dimensional vector in an embedded space via a projection function f : $V \to {{\Re }^{d}}$. For directed networks, since each node can take the role of either a source node or a target node in a relation-specific edge, we represent each node ${{v}_{i}}$ using two vector representations: a source vector ${{\vec{u}}_{i}}\in {{\Re }^{d}}$, a target vector ${{{\vec{u}}'}_{i}}\in {{\Re }^{d}}$. Also, we introduce ${{\vec{u}}_{{{r}_{s}}}}$ as the vector representation of relation ${{r}_{s}}$, as shown in the Fig.\ref{fig:directedEdge}.

Given a node $v_i$, we first define the probability that the node links to $v_j$ via a relation $r_s$, when compared with how $v_i$ is related to other nodes via its outgoing edges, denoted as
\begin{eqnarray}
\label{eq:p_out}
{{p_{out}}}( v_j, r_s | v_i )=\frac{\exp ( \vec{u}{{_{j}^{'}}^{T}}f( {{{\vec{u}}}_{i}}, {{{\vec{u}}}_{{{r}_{s}}}} ) )}{\sum\limits_{( {{v}_{i}},{{r}_{p}},{{v}_{x}} )\in E_{out}^{v_i}}{\exp ( \vec{u}{{_{x}^{'}}^{T}}f( {{{\vec{u}}}_{i}},{{{\vec{u}}}_{{{r}_{p}}}} ) )}} 
\label{multi}
\end{eqnarray}

where the source vector ${{{\vec{u}}}_{i}}$, the target vector $\vec{u}{{_{j}^{'}}}$ and the relation vector ${{{\vec{u}}}_{{{r}_{s}}}}$ for the directed edge $(v_i, r_s, v_j)$ are related by function $f(\vec{u}_i, \vec{u}_{r_s})$. The subset of $E$, $E_{out}^{v_i}$, means all the edges of which the source node are $v_i$. Note that the fuction $f(\vec{u}_i, \vec{u}_{r_s})$ is used to bridge between relations and nodes to obtain the probability compared with LINE, instead of enforcing the hard constraint as in trans-family. Likewise, the probability that the node $v_i$ is linked by $v_j$ via a relation $r_s$, when compared with how $v_i$ is related to other nodes via its input edges, denotes as:
\begin{eqnarray}
\label{eq:p_in}
{{p_{in}}}( v_j, r_s | v_i )=\frac{\exp ( \vec{u}{{_{i}^{'}}^{T}}f( {{{\vec{u}}}_{j}}, {{{\vec{u}}}_{{{r}_{s}}}} ) )}{\sum\limits_{( {{v}_{x}},{{r}_{p}},{{v}_{i}} )\in E_{in}^{v_i}}{\exp ( \vec{u}{{_{i}^{'}}^{T}}f( {{{\vec{u}}}_{x}},{{{\vec{u}}}_{{{r}_{p}}}} ) )}} 
\end{eqnarray}

Furthermore, to characterize the parallelogram structures, we take into account different possible directions of the relation edges so that three distinct non-isomorphic local connectivity structures are considered for each node in a parallelogram, as shown in Fig.\ref{fig:cases}. For the three cases, we define the corresponding probability distributions as follow:

\textbf{Case 1 (Fig.\ref{fig:case1})}: As the out-degree of $v_i$ is 2 and the in-degree of $v_i$ is 0, $p_1$ is defined as the probability that $v_i$ will ``contribute'' to such a situation, given as
\begin{equation}
\begin{split}
&p_1(v_j^{r_s},v_k^{r_t}|v_i)=\\
& \frac{\exp ( \vec{u}{{_{j}^{'}}^{T}}f( {{{\vec{u}}}_{i}}, {{{\vec{u}}}_{{{r}_{s}}}} )+\vec{u}{{_{k}^{'}}^{T}}f( {{{\vec{u}}}_{i}}, {{{\vec{u}}}_{{{r}_{t}}}} ) )}{\sum\limits_{\begin{smallmatrix}
 ( {{v}_{i}},{{r}_{p}},{{v}_{x}} )\in E_{out}^{v_i} \\
 \wedge ( {{v}_{i}},{{r}_{q}},{{v}_{y}} )\in E_{out}^{v_i}\\
 \wedge ( {{v}_{i}},{{r}_{p}},{{v}_{x}} )\neq ( {{v}_{i}},{{r}_{q}},{{v}_{y}} )
\end{smallmatrix}}{\!\exp ( \vec{u}{{_{x}^{'}}^{T}} f( {{{\vec{u}}}_{i}}, {{{\vec{u}}}_{{{r}_{p}}}} )+\vec{u}{{_{y}^{'}}^{T}}f( {{{\vec{u}}}_{i}}, {{{\vec{u}}}_{{{r}_{q}}}} ) )}}
\end{split}
\label{eq:case1}
\end{equation}

We utilize $v_i^{r_s}$ as a neater representation of the pair of $(v_i, r_s) $ in the sequel.

\textbf{Case 2 (Fig.\ref{fig:case2})}: As the out-degree of $v_i$ is 1 and the in-degree of $v_i$ is 1, $p_2$ is defined as:
\begin{equation}
\begin{split}
&p_{2}( v_{j}^{{{r}_{s}}},v_{k}^{{{r}_{t}}}|{{v}_{i}} )=\\
&\frac{\exp ( \vec{u}{{_{i}^{'}}^{T}}f( {{{\vec{u}}}_{j}}, {{{\vec{u}}}_{{{r}_{s}}}} )+\vec{u}{{_{k}^{'}}^{T}} f( {{{\vec{u}}}_{i}},{{{\vec{u}}}_{{{r}_{t}}}} ) )}{\sum\limits_{\begin{smallmatrix}
 ( {{v}_{x}},{{r}_{p}},{{v}_{i}} )\in E_{in}^{v_i} \\
 \wedge ( {{v}_{i}},{{r}_{q}},{{v}_{y}} )\in E_{out}^{v_i}
\end{smallmatrix}}{\exp ( \vec{u}{{_{i}^{'}}^{T}} f( {{{\vec{u}}}_{x}},{{{\vec{u}}}_{{{r}_{p}}}} )+\vec{u}{{_{y}^{'}}^{T}} f( {{{\vec{u}}}_{i}},{{{\vec{u}}}_{{{r}_{q}}}} ) )}}
\end{split}
\label{eq:case2}
\end{equation}

\textbf{Case 3 (Fig.\ref{fig:case3})}: As the out-degree of $v_i$ is 0 and the in-degree of $v_i$ is 2, $p_3$ is defined as:
\begin{equation}
\label{eq5}
\begin{split}
&p_{3}( v_{j}^{{{r}_{s}}},v_{k}^{{{r}_{t}}}|{{v}_{i}} )=\\
&\frac{\exp ( \vec{u}{{_{i}^{'}}^{T}} f( {{{\vec{u}}}_{j}},{{{\vec{u}}}_{{{r}_{s}}}} )+\vec{u}{{_{i}^{'}}^{T}} f( {{{\vec{u}}}_{k}},{{{\vec{u}}}_{{{r}_{t}}}} ) )}{\sum\limits_{\begin{smallmatrix}
 ( {{v}_{x}},{{r}_{p}},{{v}_{i}} )\in E_{in}^{v_i} \\
 \wedge ( {{v}_{y}},{{r}_{q}},{{v}_{i}} )\in E_{in}^{v_i}\\
 \wedge ( {{v}_{x}},{{r}_{p}},{{v}_{i}} )\neq ({{v}_{y}},{{r}_{q}},{{v}_{i}})
\end{smallmatrix}}{\!\exp ( \vec{u}{{_{i}^{'}}^{T}} f( {{{\vec{u}}}_{x}},{{{\vec{u}}}_{{{r}_{p}}}} )+\vec{u}{{_{i}^{'}}^{T}} f( {{{\vec{u}}}_{y}},{{{\vec{u}}}_{{{r}_{q}}}} ) )}}
\end{split}
\end{equation}

To preserve the three parallelogram structures, we minimize the KL-divergence of $p_{1}$, $p_{2}$, $p_{3}$ and their empirical distributions over all the nodes. The empirical distributions ${{\hat{p}}_{1}}$, ${{\hat{p}}_{2}}$ and ${{\hat{p}}_{3}}$ are defined as ${{{\omega }_{ij}}*{{\omega }_{ik}}}/({d_{out}^{i}*d_{out}^{i}})$, ${{{\omega }_{ji}}*{{\omega }_{ik}}}/({d_{in}^{i}*d_{out}^{i}})$ and ${{{\omega }_{ji}}*{{\omega }_{ki}}}/({d_{in}^{i}*d_{in}^{i}})$ respectively, where ${{\omega }_{ij}}$ denotes the weight\footnote{The weight indicates the strength of a labeled edge. In multi-relational social networks, the weight of a friendship relation between two users can be defined using the retweet frequency.} of edge $( {{v}_{i}},{{v}_{j}} )$, $d_{out}^{i}=\sum_{k\in{N_{out}^{v_i}}}w_{ik}$ and $d_{in}^{i}=\sum_{k\in{N_{in}^{v_i}}} w_{ki}$, $N_{out}^{v_i}$ and $N_{in}^{v_i}$ are the sets of out-neighbors and in-neighbors of $v_i$ respectively. As the importance of the nodes in the network may be different, we introduce $\lambda_i$ to represent the importance of $v_i$ in the network. In this paper, we set $\lambda_i$ according to its degree. Therefore, the objective function is defined as:
\begin{equation}
\label{eq15}
\begin{split}
O=\sum\limits_{i\in V}{{{\lambda }_{i}}KL\left( \hat{p}\left( \cdot |{{v}_{i}} \right)||p\left( \cdot |{{v}_{i}} \right) \right)}
\end{split}
\end{equation}
Then we set ${{\lambda }_{i}}$ to be $d_{out}^{i}*d_{out}^{i}$, $d_{in}^{i}*d_{out}^{i}$ and $d_{in}^{i}*d_{in}^{i}$ respectively, the corresponding objective function becomes:

\begin{equation}
\label{eq6}
\begin{split}
{{O}_{1}}=-\!\sum\limits_{\begin{smallmatrix}
 ( {{v}_{i}},{{r}_{s}},{{v}_{j}} )\in E \\
 \wedge ( {{v}_{i}},{{r}_{t}},{{v}_{k}} )\in E\\
 \wedge ( {{v}_{i}},{{r}_{s}},{{v}_{j}} )\neq ( {{v}_{i}},{{r}_{t}},{{v}_{k}} )
\end{smallmatrix}}{\!{\!{\!\omega }_{ij}}*{{\omega }_{ik}}*}\log {{p}_{1}}( v_{_{j}}^{{{r}_{s}}},v_{k}^{{{r}_{t}}}|{{v}_{i}})
\end{split}
\end{equation}

\begin{equation}
\label{eq7}
\begin{split}
{{O}_{2}}=-\sum\limits_{\begin{smallmatrix}
 ( {{v}_{j}},{{r}_{s}},{{v}_{i}} )\in E \\
 \wedge ( {{v}_{i}},{{r}_{t}},{{v}_{k}} )\in E
\end{smallmatrix}}{{{\omega }_{ji}}*{{\omega }_{ik}}*\log {{p}_{2}}( v_{_{j}}^{{{r}_{s}}},v_{k}^{{{r}_{t}}}|{{v}_{i}} )}
\end{split}
\end{equation}

\begin{equation}
\label{eq8}
\begin{split}
{{O}_{3}}=-\!\sum\limits_{\begin{smallmatrix}
 ( {{v}_{j}},{{r}_{s}},{{v}_{i}} )\in E \\
 \wedge ( {{v}_{k}},{{r}_{t}},{{v}_{i}} )\in E\\
 \wedge ( {{v}_{j}},{{r}_{s}},{{v}_{i}} )\neq ( {{v}_{k}},{{r}_{t}},{{v}_{i}} )
\end{smallmatrix}}\!{\!{\!{\!\omega }_{ji}}*{{\omega }_{ki}}*\log {{p}_{3}}( v_{_{j}}^{{{r}_{s}}},v_{k}^{{{r}_{t}}}|{{v}_{i}} )}
\end{split}
\end{equation}

Then, the source and target vector representations for each node, i.e., ${{\{ {{{\vec{u}}}_{i}} \}}_{i=1\ldots |V|}}$, ${{\{ \vec{u}_{i}^{'} \}}_{i=1\ldots |V|}}$ and the relation representation for each relation type, i.e., ${{\{ {{{\vec{u}}}_{{{r}_{i}}}} \}}_{i=1\ldots |R|}}$ can be obtained by minimizing the combined objective function $O={{O}_{1}}+{{O}_{2}}+{{O}_{3}}$ where $O_1$, $O_2$ and $O_3$ collaboratively help retain parallelogram structures as much as possible. In fact, the triangular structures are also implicitly preserved at the same time under such design.

\subsection{Model Inference}
The stochastic gradient descent is adopted to learn the vector representations of the multi-relational network. For example, to update the source vector of node $v_i$, the gradient w.r.t. ${{\vec{u}}_{i}}$ is computed as:
\begin{equation}
\label{eq9}
\begin{split}
&\frac{\partial O}{\partial {{{\vec{u}}}_{i}}}={{\omega }_{ij}}*{{\omega }_{ik}}*\frac{\partial \log {{p}_{1}}( v_{_{j}}^{{{r}_{s}}},v_{k}^{{{r}_{t}}}|{{v}_{i}} ) }{\partial {{{\vec{u}}}_{i}}}+\\
&{{\omega }_{ji}}({{\omega }_{ik}}*\frac{\partial \log {{p}_{2}}( v_{_{j}}^{{{r}_{s}}},v_{k}^{{{r}_{t}}}|{{v}_{i}} )}{\partial {{{\vec{u}}}_{i}}}+{{\omega }_{ki}}*\frac{\partial \log {{p}_{3}}( v_{_{j}}^{{{r}_{s}}},v_{k}^{{{r}_{t}}}|{{v}_{i}} )}{\partial {{{\vec{u}}}_{i}}})
\end{split}
\end{equation}
To reduce the computational cost of calculating the summation over the entire set of nodes when addressing the conditional probability $p_1$, $p_2$ and $p_3$, we utilize the negative sampling approach \cite{DBLP:conf/nips/MikolovSCCD13} which has been widely adopted, e.g., \cite{DBLP:conf/www/TangQWZYM15}, \cite{DBLP:journals/corr/GoldbergL14}. Negative sampling basically transforms the computationally expensive learning problem into a binary classification proxy problem that uses the same parameters but requires the statistics much easier to compute. The equivalent counterparts of the objective function Eq.(\ref{eq9}) can then be derived, given as:

\begin{equation}
\label{eq10}
\begin{split}
&\log p_{1}( v_{_{j}}^{{{r}_{s}}},v_{k}^{{{r}_{t}}}|{{v}_{i}} )\propto \log\! \sigma ( \vec{u}{{_{j}^{'}}^{T}}f( {{{\vec{u}}}_{i}},{{{\vec{u}}}_{{{r}_{s}}}} )+\vec{u}{{_{k}^{'}}^{T}}f( {{{\vec{u}}}_{i}}, {{{\vec{u}}}_{{{r}_{t}}}} ) )\\
&+\!\sum\limits_{m=1}^{K}{{E}_{\!\substack{{{{v}_{n}}\sim {{P}_{n(v)}}}\\{{{{r}_{l}}\sim {{P}_{l(r)}}}}}}\log \!\sigma (-\vec{u}{{_{j}^{'}}^{T}} f( {{{\vec{u}}}_{i}},{{{\vec{u}}}_{{{r}_{s}}}} )-\vec{u}{{_{n}^{'}}^{T}}f( {{{\vec{u}}}_{i}},{{{\vec{u}}}_{{{r}_{l}}}} )  ) }
\end{split}
\end{equation}

\begin{equation}
\label{eq11}
\begin{split}
&\log p_{2}( v_{_{j}}^{{{r}_{s}}},v_{k}^{{{r}_{t}}}|{{v}_{i}} )\propto \log \sigma (\vec{u}{{_{i}^{'}}^{T}}f( {{{\vec{u}}}_{j}},{{{\vec{u}}}_{{{r}_{s}}}} )+\vec{u}{{_{k}^{'}}^{T}}f( {{{\vec{u}}}_{i}},{{{\vec{u}}}_{{{r}_{t}}}} ) )\\
&+\sum\limits_{m=1}^{K}{{E}_{\substack{{{{v}_{n}}\sim {{P}_{n(v)}}}\\{{{{r}_{l}}\sim {{P}_{l(r)}}}}}} \log \sigma (-\vec{u}{{_{i}^{'}}^{T}}f( {{{\vec{u}}}_{j}},{{{\vec{u}}}_{{{r}_{s}}}} )-\vec{u}{{_{n}^{'}}^{T}}f( {{{\vec{u}}}_{i}},{{{\vec{u}}}_{{{r}_{l}}}} )) }
\end{split}
\end{equation}

\begin{equation}
\label{eq12}
\begin{split}
&\log p_{3}( v_{_{j}}^{{{r}_{s}}},v_{k}^{{{r}_{t}}}|{{v}_{i}} )\propto \log \sigma ( \vec{u}{{_{i}^{'}}^{T}}f( {{{\vec{u}}}_{j}},{{{\vec{u}}}_{{{r}_{s}}}} )+\vec{u}{{_{i}^{'}}^{T}}f( {{{\vec{u}}}_{k}},{{{\vec{u}}}_{{{r}_{t}}}} ) )\\
&+\sum\limits_{m=1}^{K}{{E}_{\substack{{{{v}_{n}}\sim {{P}_{n(v)}}}\\{{{{r}_{l}}\sim {{P}_{l(r)}}}}}}\log \sigma ( - \vec{u}{{_{i}^{'}}^{T}}f( {{{\vec{u}}}_{j}},{{{\vec{u}}}_{{{r}_{s}}}} )-\vec{u}{{_{i}^{'}}^{T}}f( {{{\vec{u}}}_{n}},{{{\vec{u}}}_{{{r}_{l}}}} ) )  }
\end{split}
\end{equation}
Each of the first terms of Eqs.(\ref{eq10}-\ref{eq12}) models the observed local structures (positive samples), while each of the second terms models the way the negative samples drawn from the noise distribution (we adopt uniform distribution in this paper). $\sigma ( x ){ }={ }1/( 1{ }+{ }exp( -x ) )$ denotes the sigmoid function. ${{v}_{n}}$ and $r_l$ denote the negative samples for nodes and relation edges drawn from a uniform distribution where $v_i$, $r_l$ and ${{v}_{n}}$ cannot constitute the fact triplet, and $K$ is the number of the negative samples. 
\subsubsection{Bridging by addition}
The bridging function $f(\vec{u}_i,\vec{u}_{r_s})$ can be simply facilitated with addition:
\begin{equation}
f(\vec{u}_i,\vec{u}_{r_s}) = \vec{u}_i + \vec{u}_{r_s}.
\end{equation}
And the proposed multi-relational network embedding (MNE) model with such bridging function will be referred to as ${\!M\!N\!E}^+$ in the sequel. Then the partial derivative of Eq.(\ref{eq9}), by replacing $\log p_1(v_j^{r_s},v_k^{r_t}|v_i)$, $\log p_2(v_j^{r_s},v_k^{r_t}|v_i)$, $\log p_3(v_j^{r_s},v_k^{r_t}|v_i)$ with Eq.(\ref{eq10}), Eq.(\ref{eq11}) and Eq.(\ref{eq12}) respectively, can be rewritten as:
\begin{equation}
\label{updateui}
\begin{split}
&\frac{\partial {O}}{\partial {{{\vec{u}}}_{i}}}=\\
&{{\omega }_{ij}}*{{\omega }_{ik}}*\Big([1-\sigma (\vec{u}{{_{j}^{'}}^{T}}( {{\vec{u}}_{i}}\!+\!{{\vec{u}}_{{r}_{s}}} )+\vec{u}{{_{k}^{'}}^{T}}( {{{\vec{u}}}_{i}}\!+\!{{\vec{u}}_{{r}_{t}}} ) ) ]*(\vec{u}_{j}^{'}\!+\!\vec{u}_{k}^{'} ) \\
&-\sum_{m=1}^{K}{E}_{\substack{{{{v}_{n}}\sim {{P}_{n(v)}}}\\{{{{r}_{l}}\sim {{P}_{l(r)}}}}}}\sigma ( \vec{u}{{_{j}^{'}}^{T}}( {{\vec{u}}_{i}}\!+\!{{\vec{u}}_{{r}_{s}}} )\!+\!\vec{u}{{_{n}^{'}}^{T}}( {{{\vec{u}}}_{i}}\!+\!{{\vec{u}}_{{r}_{l}}} ) ) (\vec{u}_{j}^{'}\!+\!\vec{u}_{n}^{'} )\Big)\\
&+{{\omega }_{ji}}*{{\omega }_{ik}}*\Big([1-\sigma ( \vec{u}{{_{i}^{'}}^{T}} ( {{\vec{u}}_{j}}+{{\vec{u}}_{{r}_{s}}} )+\vec{u}{{_{k}^{'}}^{T}}( {{{\vec{u}}}_{i}}+{{\vec{u}}_{{r}_{t}}} ) ) ]*\vec{u}_{k}^{'} \\
&-\sum_{m=1}^{K}{E}_{\substack{{{{v}_{n}}\sim {{P}_{n(v)}}}\\{{{{r}_{l}}\sim {{P}_{l(r)}}}}}}\sigma ( \vec{u}{{_{i}^{'}}^{T}}( {{\vec{u}}_{j}}+{{\vec{u}}_{{r}_{s}}} )+\vec{u}{{_{n}^{'}}^{T}}( {{{\vec{u}}}_{i}}+{{\vec{u}}_{{r}_{l}}} ) ) *\vec{u}_{n}^{'} \Big)
\end{split}
\end{equation}

\begin{equation}
\label{updateui1}
\begin{split}
&\frac{\partial {O}}{\partial {{{\vec{u}^{'}}}_{i}}}={{\omega }_{ji}}*{{\omega }_{ik}}*\Big([1-\sigma (\vec{u}{{_{i}^{'}}^{T}}( {{\vec{u}}_{j}}+{{\vec{u}}_{{r}_{s}}} )+\vec{u}{{_{k}^{'}}^{T}}( {{{\vec{u}}}_{i}}+{{\vec{u}}_{{r}_{t}}} ) )]  \\
&-\sum_{m=1}^{K}{E}_{\substack{{{{v}_{n}}\sim {{P}_{n(v)}}}\\{{{{r}_{l}}\sim {{P}_{l(r)}}}}}}\sigma ( \vec{u}{{_{i}^{'}}^{T}}( {{\vec{u}}_{j}}\!+\!{{\vec{u}}_{{r}_{s}}} )\!+\!\vec{u}{{_{n}^{'}}^{T}}( {{{\vec{u}}}_{i}}\!+\!{{\vec{u}}_{{r}_{l}}} ) )\Big)*(\vec{u}_{j}\!+\!\vec{u}_{r_s} )\\
&+\!{{\omega }_{ji}}*{{\omega }_{ki}}*\Big([1-\sigma ( \vec{u}{{_{i}^{'}}^{T}} ( {{\vec{u}}_{j}}+{{\vec{u}}_{{r}_{s}}} )+\vec{u}{{_{i}^{'}}^{T}}( {{{\vec{u}}}_{k}}+{{\vec{u}}_{{r}_{t}}} ) ) ]\\
&*(\vec{u}_{j}+\vec{u}_{{r}_{s}}+\vec{u}_{k}+\vec{u}_{{r}_{t}})\\
&-\!\sum_{m=1}^{K}{E}_{\substack{{{{v}_{n}}\sim {{P}_{n(v)}}}\\{{{{r}_{l}}\sim {{P}_{l(r)}}}}}}\sigma ( \vec{u}{{_{i}^{'}}^{T}}( {{\vec{u}}_{j}}+{{\vec{u}}_{{r}_{s}}} )+\vec{u}{{_{i}^{'}}^{T}}( {{{\vec{u}}}_{n}}+{{\vec{u}}_{{r}_{l}}} ) ) \\
&*(\vec{u}_{j}+\vec{u}_{r_s}+\vec{u}_{n}+\vec{u}_{r_l} )\Big)
\end{split}
\end{equation}

\begin{equation}
\label{updateuj}
\begin{split}
&\frac{\partial {O}}{\partial {{{\vec{u}}}_{j}}}={{\omega }_{ji}}*{{\omega }_{ik}}*\Big([1-\sigma (\vec{u}{{_{i}^{'}}^{T}}( {{\vec{u}}_{j}}+{{\vec{u}}_{{r}_{s}}} )+\vec{u}{{_{k}^{'}}^{T}}( {{{\vec{u}}}_{i}}+{{\vec{u}}_{{r}_{t}}} ) )  ]\\
&-\sum_{m=1}^{K}{E}_{\substack{{{{v}_{n}}\sim {{P}_{n(v)}}}\\{{{{r}_{l}}\sim {{P}_{l(r)}}}}}}\sigma ( \vec{u}{{_{i}^{'}}^{T}}( {{\vec{u}}_{j}}+{{\vec{u}}_{{r}_{s}}} )+\vec{u}{{_{n}^{'}}^{T}}( {{{\vec{u}}}_{i}}+{{\vec{u}}_{{r}_{l}}} ) ) \Big)*\vec{u}_{i}^{'}\\
&+{{\omega }_{ji}}*{{\omega }_{ki}}*\Big([1-\sigma ( \vec{u}{{_{i}^{'}}^{T}} ( {{\vec{u}}_{j}}+{{\vec{u}}_{{r}_{s}}} )+\vec{u}{{_{i}^{'}}^{T}}( {{{\vec{u}}}_{k}}+{{\vec{u}}_{{r}_{t}}} ) ) ]\\
&-\sigma ( \vec{u}{{_{i}^{'}}^{T}}( {{\vec{u}}_{j}}+{{\vec{u}}_{{r}_{s}}} )+\vec{u}{{_{i}^{'}}^{T}}( {{{\vec{u}}}_{n}}+{{\vec{u}}_{{r}_{l}}} ) ) \Big)*\vec{u}_{i}^{'}
\end{split}
\end{equation}

\begin{equation}
\label{updateuj1}
\begin{split}
&\frac{\partial {O}}{\partial {{{\vec{u}_{j}}^{'}}}}={{\omega }_{ij}}*{{\omega }_{ik}}*\Big( [1\!-\!\sigma (\vec{u}{{_{j}^{'}}^{T}}( {{\vec{u}}_{i}}\!+\!{{\vec{u}}_{{r}_{s}}} )\!+\!\vec{u}{{_{k}^{'}}^{T}}( {{{\vec{u}}}_{i}}\!+\!{{\vec{u}}_{{r}_{t}}} ) ) ] \\
&-\!\sum_{m=1}^{K}{E}_{\substack{{{{v}_{n}}\sim {{P}_{n(v)}}}\\{{{{r}_{l}}\sim {{P}_{l(r)}}}}}}\sigma ( \vec{u}{{_{j}^{'}}^{T}}( {{\vec{u}}_{i}}\!+\!{{\vec{u}}_{{r}_{s}}} )\!+\!\vec{u}{{_{n}^{'}}^{T}}( {{{\vec{u}}}_{i}}\!+\!{{\vec{u}}_{{r}_{l}}} ) ) \Big)\left(\vec{u}_{i}\!+\!\vec{u}_{r_s} \right)
\end{split}
\end{equation}

\begin{equation}
\label{updateuk}
\begin{split}
&\frac{\partial {O}}{\partial {{{\vec{u}}}_{k}}}=\Big(1-\sigma (\vec{u}{{_{i}^{'}}^{T}}( {{\vec{u}}_{j}}+{{\vec{u}}_{{r}_{s}}} )+\vec{u}{{_{i}^{'}}^{T}}( {{{\vec{u}}}_{k}}+{{\vec{u}}_{{r}_{t}}} ) ) \Big)* {{\vec{u}_{i}}^{'}}
\end{split}
\end{equation}

\begin{equation}
\label{updateuk1}
\begin{split}
&\!\frac{\!\partial {O}}{\!\partial {{{\vec{u}_{k}}^{'}}}}\!=\!{{\omega }_{ij}}\!*\!{{\omega }_{ik}}*\Big(\![1\!-\!\sigma\! (\vec{u}{{_{j}^{'}}\!^{T}}\!({{\vec{u}}_{i}}\!+{{\vec{u}}_{{r}_{s}}} )\!+\vec{u}{{_{k}^{'}}\!^{T}}\!( {{{\vec{u}}}_{i}}\!+\!{{\vec{u}}_{{r}_{t}}} ) ) ]\!*\!(\vec{u}_{i}\!+\vec{u}_{r_t} )\Big)\\
&\!+\!{{\omega }_{ji}}\!*\!{{\omega }_{ik}}*\Big([1\!-\!\sigma(\vec{u}{{_{i}^{'}}\!^{T}}({{\vec{u}}_{j}}\!+\!{{\vec{u}}_{{r}_{s}}})\!+\!\vec{u}{{_{k}^{'}}\!^{T}}({{{\vec{u}}}_{i}}\!+\!{{\vec{u}}_{{r}_{t}}}))]\!*\!(\vec{u}_{i}\!+\!\vec{u}_{r_t})\Big)
\end{split}
\end{equation}
\begin{equation}
\label{updateurs}
\begin{split}
&\frac{\partial {O}}{\partial {{{\vec{u}}}_{r_s}}}={{\omega }_{ij}}*{{\omega }_{ik}}*\Big([1-\sigma (\vec{u}{{_{j}^{'}}^{T}}( {{\vec{u}}_{i}}+{{\vec{u}}_{{r}_{s}}} )+\vec{u}{{_{k}^{'}}^{T}}( {{{\vec{u}}}_{i}}+{{\vec{u}}_{{r}_{t}}} ) ) ] \\
&-\sum_{m=1}^{K}{E}_{\substack{{{{v}_{n}}\sim {{P}_{n(v)}}}\\{{{{r}_{l}}\sim {{P}_{l(r)}}}}}}\sigma ( \vec{u}{{_{j}^{'}}^{T}}( {{\vec{u}}_{i}}+{{\vec{u}}_{{r}_{s}}} )+\vec{u}{{_{n}^{'}}^{T}}( {{{\vec{u}}}_{i}}+{{\vec{u}}_{{r}_{l}}} ) ) \Big)*\vec{u}_{j}^{'}\\
&+{{\omega }_{ji}}*{{\omega }_{ik}}*\Big([1-\sigma (\vec{u}{{_{i}^{'}}^{T}}( {{\vec{u}}_{j}}+{{\vec{u}}_{{r}_{s}}} )+\vec{u}{{_{k}^{'}}^{T}}( {{{\vec{u}}}_{i}}+{{\vec{u}}_{{r}_{t}}} ) ) ] \\
&-\sum_{m=1}^{K}{E}_{\substack{{{{v}_{n}}\sim {{P}_{n(v)}}}\\{{{{r}_{l}}\sim {{P}_{l(r)}}}}}}\sigma ( \vec{u}{{_{i}^{'}}^{T}}( {{\vec{u}}_{j}}+{{\vec{u}}_{{r}_{s}}} )+\vec{u}{{_{n}^{'}}^{T}}( {{{\vec{u}}}_{i}}+{{\vec{u}}_{{r}_{l}}} ) ) \Big)*\vec{{u}_{i}}^{'}\\
&+{{\omega }_{ji}}*{{\omega }_{ki}}*\Big([1-\sigma (\vec{u}{{_{i}^{'}}^{T}}( {{\vec{u}}_{j}}+{{\vec{u}}_{{r}_{s}}} )+\vec{u}{{_{i}^{'}}^{T}}( {{{\vec{u}}}_{k}}+{{\vec{u}}_{{r}_{t}}} ) ) ] \\
&-\sum_{m=1}^{K}{E}_{\substack{{{{v}_{n}}\sim {{P}_{n(v)}}}\\{{{{r}_{l}}\sim {{P}_{l(r)}}}}}}\sigma ( \vec{u}{{_{i}^{'}}^{T}}( {{\vec{u}}_{j}}+{{\vec{u}}_{{r}_{s}}} )+\vec{u}{{_{i}^{'}}^{T}}( {{{\vec{u}}}_{n}}+{{\vec{u}}_{{r}_{l}}} ) ) \Big)*\vec{{u}_{i}}^{'} \\
\end{split}
\end{equation}
\begin{equation}
\label{updateurt_add}
\begin{split}
&\frac{\partial {O}}{\partial {{{\vec{u}}}_{r_t}}}={{\omega }_{ij}}\!*\!{{\omega }_{ik}}\!*\!\Big(1-\sigma (\vec{u}{{_{j}^{'}}\!^{T}}({{\vec{u}}_{i}}\!+\!{{\vec{u}}_{{r}_{s}}})\!+\!\vec{u}{{_{k}^{'}}\!^{T}}({{{\vec{u}}}_{i}}\!+\!{{\vec{u}}_{{r}_{t}}}))\Big)\!*\!\vec{u}_{k}^{'}\!\\
&+{{\omega }_{ji}}*{{\omega }_{ik}}*\Big(1-\sigma ( \vec{u}{{_{i}^{'}}^{T}} ( {{\vec{u}}_{j}}+{{\vec{u}}_{{r}_{s}}} )+\vec{u}{{_{k}^{'}}^{T}}( {{{\vec{u}}}_{i}}+{{\vec{u}}_{{r}_{t}}} ) ) \Big)\!*\!\vec{u}_{k}^{'}\\
&+{{\omega }_{ji}}*{{\omega }_{ki}}\Big(1-\sigma ( \vec{u}{{_{i}^{'}}^{T}} ( {{\vec{u}}_{j}}+{{\vec{u}}_{{r}_{s}}} )+\vec{u}{{_{i}^{'}}^{T}}( {{{\vec{u}}}_{k}}+{{\vec{u}}_{{r}_{t}}} ) ) \Big)*\vec{u}_{i}^{'}
\end{split}
\end{equation}

With reference to Eq.(\ref{updateui}-\ref{updateurt_add}), the updating rule for the embedding vector $\vec u_i$, the target vectors $\vec{u}_i^{'}$ and relation vectors ${\vec{u}}_{{r}_{l}}$ can be obtained. 
\subsubsection{Bridging by multiplication}
Alternatively, we also come up with another form of the bridging function by adopting the product operation, given as:
\begin{equation}
f(\vec{u}_i,\vec{u}_{r_s}) = {\vec{u}_{r_s}}\cdot{\vec{u}_{r_s}}^T\cdot{\vec{u}_i}.
\end{equation}
Our proposed multi-relational network embedding by use of the above bridging function will be referred to as ${\!M\!N\!E}^*$ in the sequel. The counterparts of the partial derivative of Eq.(\ref{eq9}) for ${\!M\!N\!E}^*$ can then be derived as:

\begin{equation}
\label{updateui*}
\begin{split}
&\!\frac{\!\partial{O}}{\!\partial{{{\vec{u}}}_{i}}}\!=\!{{\omega }_{ij}}\!*\!{{\omega }_{ik}}\!*\!\Big([1\!-\!\sigma( ( \vec{u}{{_{j}^{'}}\!^{T}}\!\cdot\!{{{\vec{u}}}_{{{r}_{s}}}}\!\cdot\!{{{\vec{u}}}_{{{r}_{s}}}}\!^{T}\!\cdot\!{{{\vec{u}}}_{i}} )\!+\!(\vec{u}{{_{k}^{'}}\!^{T}}\!\cdot\!{{{\vec{u}}}_{{{r}_{t}}}}\!\cdot\!{{{\vec{u}}}_{{{r}_{t}}}}\!^{T}\!\cdot\!{{{\vec{u}}}_{i}}))] \\
&*({{{\vec{u}}}_{{{r}_{s}}}}\cdot{{{\vec{u}}}_{{{r}_{s}}}}^{T}\cdot \vec{u}{{_{j}^{'}}}+ {{{\vec{u}}}_{{{r}_{t}}}}\cdot{{{\vec{u}}}_{{{r}_{t}}}}^{T}\cdot \vec{u}{{_{k}^{'}}})\\
&-\!\sum_{m=1}^{K}{E}_{\substack{{{{v}_{n}}\sim {{P}_{n(v)}}}\\{{{{r}_{l}}\sim {{P}_{l(r)}}}}}}({{{\vec{u}}}_{{{r}_{s}}}}\cdot {{{\vec{u}}}_{{{r}_{s}}}}^{T}\cdot \vec{u}{{_{j}^{'}}}+{{{\vec{u}}}_{{{r}_{l}}}}\cdot{{{\vec{u}}}_{{{r}_{l}}}}^{T}\cdot \vec{u}{{_{n}^{'}}})\\
&*\sigma( ( \vec{u}{{_{j}^{'}}^{T}}\cdot {{{\vec{u}}}_{{{r}_{s}}}}\cdot {{{\vec{u}}}_{{{r}_{s}}}}^{T}\cdot {{{\vec{u}}}_{i}} )+( \vec{u}{{_{n}^{'}}^{T}}\cdot {{{\vec{u}}}_{{{r}_{l}}}}\cdot {{{\vec{u}}}_{{{r}_{l}}}}^{T}\cdot {{{\vec{u}}}_{i}} ))\Big)\\
&+{{\omega }_{ji}}*{{\omega }_{ik}}\!*\!\Big([1\!-\!\sigma( ( \vec{u}{{_{i}^{'}}\!^{T}}\!\cdot\!{{{\vec{u}}}_{{{r}_{s}}}}\!\cdot\!{{{\vec{u}}}_{{{r}_{s}}}}\!^{T}\!\cdot\!{{{\vec{u}}}_{j}} )\!+\!(\vec{u}{{_{k}^{'}}\!^{T}}\!\cdot\!{{{\vec{u}}}_{{{r}_{t}}}}\!\cdot\!{{{\vec{u}}}_{{{r}_{t}}}}\!^{T}\!\cdot\!{{{\vec{u}}}_{i}} ))]\\
&*({{{\vec{u}}}_{{{r}_{t}}}}\!\cdot\!{{{\vec{u}}}_{{{r}_{t}}}}\!^{T}\!\cdot\!\vec{u}{{_{k}^{'}}})-\sum_{m=1}^{K}{E}_{\substack{{{{v}_{n}}\sim {{P}_{n(v)}}}\\{{{{r}_{l}}\sim {{P}_{l(r)}}}}}}\!({{{\vec{u}}}_{{{r}_{l}}}}\!\cdot\!{{{\vec{u}}}_{{{r}_{l}}}}\!^{T}\!\cdot\!\vec{u}{{_{n}^{'}}})\!\\
&*\sigma((\vec{u}{{_{i}^{'}}\!^{T}}\!\cdot\!{{{\vec{u}}}_{{{r}_{s}}}}\!\cdot\!{{{\vec{u}}}_{{{r}_{s}}}}\!^{T}\!\cdot\!{{{\vec{u}}}_{j}})\!+\!(\vec{u}{{_{n}^{'}}\!^{T}}\!\cdot\!{{{\vec{u}}}_{{{r}_{l}}}}\!\cdot\!{{{\vec{u}}}_{{{r}_{l}}}}\!^{T}\!\cdot\!{{{\vec{u}}}_{i}} ))\Big)
\end{split}
\end{equation}
\begin{equation}
\label{updateui1*}
\begin{split}
&\frac{\partial {O}}{\partial {{{\vec{u}^{'}}}_{i}}}\!=\!{{\omega }_{ji}}*{{\omega }_{ik}}*({{{\vec{u}}}_{{{r_s}}}}\cdot {{{\vec{u}}}_{{{r}_{s}^T}}}\!\cdot\! {{{\vec{u}}}_{j}})\\
&*\Big([1-\sigma( ( \vec{u}{{_{i}^{'}}\!^{T}}\!\cdot\!{{{\vec{u}}}_{{{r}_{s}}}}\!\cdot\!{{{\vec{u}}}_{{{r}_{s}}}}\!^{T}\!\cdot\!{{{\vec{u}}}_{j}} )\!+\!( \vec{u}{{_{k}^{'}}\!^{T}}\!\cdot\!{{{\vec{u}}}_{{{r}_{t}}}}\!\cdot\!{{{\vec{u}}}_{{{r}_{t}}}}\!^{T}\!\cdot\! {{{\vec{u}}}_{i}}))]\\
&\!-\sum_{m=1}^{K}{E}_{\substack{{{{v}_{n}}\sim {{P}_{n(v)}}}\\{{{{r}_{l}}\sim {{P}_{l(r)}}}}}}\!\sigma( ( \vec{u}{{_{i}^{'}}^{T}}\!\cdot\! {{{\vec{u}}}_{{{r}_{s}}}}\!\cdot\! {{{\vec{u}}}_{{{r}_{s}}}}^{T}\!\cdot\! {{{\vec{u}}}_{j}} )\!+\!( \vec{u}{{_{n}^{'}}^{T}}\!\cdot\! {{{\vec{u}}}_{{{r}_{l}}}}\!\cdot\! {{{\vec{u}}}_{{{r}_{l}}}}^{T}\!\cdot\! {{{\vec{u}}}_{i}} ))\Big)\\
&+{{\omega }_{ji}}*{{\omega }_{ki}}*\Big((\vec{u}_{r_s}\cdot {{{\vec{u}}}_{{{r}_{s}}}^T}\cdot {{{\vec{u}}}_{j}}+ {{{\vec{u}}}_{{{r_t}}}}\cdot {{{\vec{u}}}_{{{r}_{t}}}^T}\cdot {{{\vec{u}}}_{k}} )\\
&*[1-\sigma( ( \vec{u}{{_{i}^{'}}^{T}}\cdot {{{\vec{u}}}_{{{r}_{s}}}}\cdot {{{\vec{u}}}_{{{r}_{s}}}}^{T}\cdot {{{\vec{u}}}_{j}} )+( \vec{u}{{_{i}^{'}}^{T}}\cdot {{{\vec{u}}}_{{{r}_{t}}}}\cdot {{{\vec{u}}}_{{{r}_{t}}}}^{T}\cdot {{{\vec{u}}}_{k}} ))]\\
&- \sum_{m=1}^{K}{E}_{\substack{{{{v}_{n}}\sim {{P}_{n(v)}}}\\{{{{r}_{l}}\sim {{P}_{l(r)}}}}}}({{{\vec{u}}}_{{{r}_{s}}}}\cdot {{{\vec{u}}}_{{{r}_{s}}}}^{T}\cdot {{{\vec{u}}}_{j}}+ {{{\vec{u}}}_{{{r}_{l}}}}\cdot {{{\vec{u}}}_{{{r}_{l}}}}^{T}\cdot {{{\vec{u}}}_{n}} )\\
&*\sigma( ( \vec{u}{{_{i}^{'}}^{T}}\cdot {{{\vec{u}}}_{{{r}_{s}}}}\cdot {{{\vec{u}}}_{{{r}_{s}}}}^{T}\cdot {{{\vec{u}}}_{j}} )+( \vec{u}{{_{i}^{'}}^{T}}\cdot {{{\vec{u}}}_{{{r}_{l}}}}\cdot {{{\vec{u}}}_{{{r}_{l}}}}^{T}\cdot {{{\vec{u}}}_{n}} ))\Big)
\end{split}
\end{equation}
\begin{equation}
\label{updateuj*}
\begin{split}
&\frac{\partial {O}}{\partial {{{\vec{u}}}_{j}}}={{\omega }_{ji}}*{{\omega }_{ik}}*({{{\vec{u}}}_{{{r}_{s}}}}\cdot{{{\vec{u}}}_{{{r}_{s}}}}^{T}\cdot\vec{u}{{_{i}^{'}}})\\
&*\Big([1-\sigma( ( \vec{u}{{_{i}^{'}}^{T}}\!\cdot\! {{{\vec{u}}}_{{{r}_{s}}}}\!\cdot \!{{{\vec{u}}}_{{{r}_{s}}}}\!^{T}\!\cdot\! {{{\vec{u}}}_{j}} )\!+\!( \vec{u}{{_{k}^{'}}\!^{T}}\!\cdot\! {{{\vec{u}}}_{{{r}_{t}}}}\!\cdot\! {{{\vec{u}}}_{{{r}_{t}}}}\!^{T}\!\cdot\! {{{\vec{u}}}_{i}} ))]\\
&-\sum_{m=1}^{K}{E}_{\substack{{{{v}_{n}}\sim {{P}_{n(v)}}}\\{{{{r}_{l}}\sim {{P}_{l(r)}}}}}}\sigma( ( \vec{u}{{_{i}^{'}}\!^{T}}\!\cdot\! {{{\vec{u}}}_{{{r}_{s}}}}\!\cdot\! {{{\vec{u}}}_{{{r}_{s}}}}\!^{T}\!\cdot\! {{{\vec{u}}}_{j}} )\!+\!( \vec{u}{{_{k}^{'}}\!^{T}}\!\cdot\! {{{\vec{u}}}_{{{r}_{l}}}}\!\cdot\! {{{\vec{u}}}_{{{r}_{l}}}}\!^{T}\!\cdot\! {{{\vec{u}}}_{i}} ))\Big) \\
&+{{\omega }_{ji}}*{{\omega }_{ki}}*({{{\vec{u}}}_{{{r}_{s}}}}\cdot{{{\vec{u}}}_{{{r}_{s}}}}^{T}\cdot\vec{u}{{_{i}^{'}}})\\
&*\Big([1-\sigma( ( \vec{u}{{_{i}^{'}}^{T}}\cdot {{{\vec{u}}}_{{{r}_{s}}}}\cdot {{{\vec{u}}}_{{{r}_{s}}}}^{T}\cdot {{{\vec{u}}}_{j}} )+( \vec{u}{{_{i}^{'}}^{T}}\cdot {{{\vec{u}}}_{{{r}_{t}}}}\cdot {{{\vec{u}}}_{{{r}_{t}}}}^{T}\cdot {{{\vec{u}}}_{k}} ))]\\
&-\sum_{m=1}^{K}{E}_{\substack{{{{v}_{n}}\sim {{P}_{n(v)}}}\\{{{{r}_{l}}\sim {{P}_{l(r)}}}}}}\sigma( ( \vec{u}{{_{i}^{'}}^{T}}\cdot {{{\vec{u}}}_{{{r}_{s}}}}\cdot {{{\vec{u}}}_{{{r}_{s}}}}^{T}\cdot {{{\vec{u}}}_{j}} )+( \vec{u}{{_{i}^{'}}^{T}}\cdot {{{\vec{u}}}_{{{r}_{l}}}}\cdot {{{\vec{u}}}_{{{r}_{l}}}}^{T}\cdot {{{\vec{u}}}_{n}} ))\Big)
\end{split}
\end{equation}
\begin{equation}
\label{updateuj1*}
\begin{split}
&\frac{\partial {O}}{\partial {{{\vec{u}_{j}}^{'}}}}={{\omega }_{ij}}*{{\omega }_{ik}}*({{{\vec{u}}}_{{{r}_{s}}}}\cdot {{{\vec{u}}}_{{{r}_{s}}}}^{T}\cdot {{{\vec{u}}}_{i}})\\
&*\Big([1-\sigma( ( \vec{u}{{_{j}^{'}}^{T}}\cdot {{{\vec{u}}}_{{{r}_{s}}}}\cdot {{{\vec{u}}}_{{{r}_{s}}}}^{T}\cdot {{{\vec{u}}}_{i}} )+( \vec{u}{{_{k}^{'}}^{T}}\cdot {{{\vec{u}}}_{{{r}_{t}}}}\cdot {{{\vec{u}}}_{{{r}_{t}}}}^{T}\cdot {{{\vec{u}}}_{i}} ))]\\
&-\sum_{m=1}^{K}{E}_{\substack{{{{v}_{n}}\sim {{P}_{n(v)}}}\\{{{{r}_{l}}\sim {{P}_{l(r)}}}}}}\sigma( ( \vec{u}{{_{j}^{'}}^{T}}\cdot {{{\vec{u}}}_{{{r}_{s}}}}\cdot {{{\vec{u}}}_{{{r}_{s}}}}^{T}\cdot {{{\vec{u}}}_{i}} )+( \vec{u}{{_{n}^{'}}^{T}}\cdot {{{\vec{u}}}_{{{r}_{l}}}}\cdot {{{\vec{u}}}_{{{r}_{l}}}}^{T}\cdot {{{\vec{u}}}_{i}} ))\Big)\\
\end{split}
\end{equation}
\begin{equation}
\label{updateuk*}
\begin{split}
&\frac{\partial {O}}{\partial {{{\vec{u}}}_{k}}}={{\omega }_{ji}}*{{\omega }_{ki}}*({{{\vec{u}}}_{{{r}_{t}}}}\cdot {{{\vec{u}}}_{{{r}_{t}}}}^{T}\cdot {{{\vec{u}}}_{i}})\\
&*\Big(1-\sigma( ( \vec{u}{{_{i}^{'}}^{T}}\cdot {{{\vec{u}}}_{{{r}_{s}}}}\cdot {{{\vec{u}}}_{{{r}_{s}}}}^{T}\cdot {{{\vec{u}}}_{j}} )+( \vec{u}{{_{i}^{'}}^{T}}\cdot {{{\vec{u}}}_{{{r}_{t}}}}\cdot {{{\vec{u}}}_{{{r}_{t}}}}^{T}\cdot {{{\vec{u}}}_{k}} ))\Big)\\
&
\end{split}
\end{equation}
\begin{equation}
\label{updateuk1*}
\begin{split}
&\frac{\partial {O}}{\partial {{{\vec{u}_{k}}^{'}}}}={{\omega }_{ij}}*{{\omega }_{ik}}*(\cdot {{{\vec{u}}}_{{{r}_{t}}}}\cdot {{{\vec{u}}}_{{{r}_{t}}}}^{T}\cdot {{{\vec{u}}}_{i}})\\
&*\Big(1-\sigma( ( \vec{u}{{_{j}^{'}}^{T}}\cdot {{{\vec{u}}}_{{{r}_{s}}}}\cdot {{{\vec{u}}}_{{{r}_{s}}}}^{T}\cdot {{{\vec{u}}}_{i}} )+( \vec{u}{{_{k}^{'}}^{T}}\cdot {{{\vec{u}}}_{{{r}_{t}}}}\cdot {{{\vec{u}}}_{{{r}_{t}}}}^{T}\cdot {{{\vec{u}}}_{i}} ))\Big)\\
&+{{\omega }_{ji}}*{{\omega }_{ik}}*(\cdot {{{\vec{u}}}_{{{r}_{t}}}}\cdot {{{\vec{u}}}_{{{r}_{t}}}}^{T}\cdot {{{\vec{u}}}_{i}})\\
&*\Big(1-\sigma( ( \vec{u}{{_{i}^{'}}^{T}}\cdot {{{\vec{u}}}_{{{r}_{s}}}}\cdot {{{\vec{u}}}_{{{r}_{s}}}}^{T}\cdot {{{\vec{u}}}_{j}} )+( \vec{u}{{_{k}^{'}}^{T}}\cdot {{{\vec{u}}}_{{{r}_{t}}}}\cdot {{{\vec{u}}}_{{{r}_{t}}}}^{T}\cdot {{{\vec{u}}}_{i}} ))\Big)
\end{split}
\end{equation}
\begin{equation}
\label{updateurs*}
\begin{split}
&\frac{\partial {O}}{\partial {{{\vec{u}}}_{r_s}}}={{\omega }_{ij}}*{{\omega }_{ik}}*(2\vec{u}{{_{j}^{'}}^{T}}\cdot {{{\vec{u}}}_{{{r}_{s}}}}\cdot {{{\vec{u}}}_{i}})\\
&*\Big([1-\sigma( ( \vec{u}{{_{j}^{'}}^{T}}\cdot {{{\vec{u}}}_{{{r}_{s}}}}\cdot {{{\vec{u}}}_{{{r}_{s}}}}^{T}\cdot {{{\vec{u}}}_{i}} )+( \vec{u}{{_{k}^{'}}^{T}}\cdot {{{\vec{u}}}_{{{r}_{t}}}}\cdot {{{\vec{u}}}_{{{r}_{t}}}}^{T}\cdot {{{\vec{u}}}_{i}} ))]\\
&-\sum_{m=1}^{K}{E}_{\substack{{{{v}_{n}}\sim {{P}_{n(v)}}}\\{{{{r}_{l}}\sim {{P}_{l(r)}}}}}}\sigma( ( \vec{u}{{_{j}^{'}}^{T}}\!\cdot\! {{{\vec{u}}}_{{{r}_{s}}}}\!\cdot\! {{{\vec{u}}}_{{{r}_{s}}}}^{T}\!\cdot\! {{{\vec{u}}}_{i}} )\!+\!( \vec{u}{{_{n}^{'}}^{T}}\!\cdot\! {{{\vec{u}}}_{{{r}_{l}}}}\!\cdot\! {{{\vec{u}}}_{{{r}_{l}}}}^{T}\!\cdot\! {{{\vec{u}}}_{i}} ))\Big)\\
&+{{\omega }_{ji}}*{{\omega }_{ik}}*(2\vec{u}{{_{i}^{'}}^{T}}\cdot {{{\vec{u}}}_{{{r}_{s}}}}\cdot {{{\vec{u}}}_{j}})\\
&*\Big([1-\sigma( ( \vec{u}{{_{i}^{'}}^{T}}\cdot {{{\vec{u}}}_{{{r}_{s}}}}\cdot {{{\vec{u}}}_{{{r}_{s}}}}^{T}\cdot {{{\vec{u}}}_{j}} )+( \vec{u}{{_{k}^{'}}^{T}}\cdot {{{\vec{u}}}_{{{r}_{t}}}}\cdot {{{\vec{u}}}_{{{r}_{t}}}}^{T}\cdot {{{\vec{u}}}_{i}} ))]\\
&-\sum_{m=1}^{K}{E}_{\substack{{{{v}_{n}}\sim {{P}_{n(v)}}}\\{{{{r}_{l}}\sim {{P}_{l(r)}}}}}}\sigma( ( \vec{u}{{_{i}^{'}}^{T}}\!\cdot\! {{{\vec{u}}}_{{{r}_{s}}}}\!\cdot\! {{{\vec{u}}}_{{{r}_{s}}}}^{T}\!\cdot\! {{{\vec{u}}}_{j}} )\!+\!( \vec{u}{{_{n}^{'}}^{T}}\!\cdot\! {{{\vec{u}}}_{{{r}_{l}}}}\!\cdot\! {{{\vec{u}}}_{{{r}_{l}}}}^{T}\!\cdot\! {{{\vec{u}}}_{i}} ))\Big)\\
&+{{\omega }_{ji}}*{{\omega }_{ki}}*(2\vec{u}{{_{i}^{'}}^{T}}\cdot {{{\vec{u}}}_{{{r}_{s}}}}\cdot {{{\vec{u}}}_{j}})\\
&*\Big([1-\sigma( ( \vec{u}{{_{i}^{'}}^{T}}\cdot {{{\vec{u}}}_{{{r}_{s}}}}\cdot {{{\vec{u}}}_{{{r}_{s}}}}^{T}\cdot {{{\vec{u}}}_{j}} )+( \vec{u}{{_{i}^{'}}^{T}}\cdot {{{\vec{u}}}_{{{r}_{t}}}}\cdot {{{\vec{u}}}_{{{r}_{t}}}}^{T}\cdot {{{\vec{u}}}_{k}} ))]\\
&-\sum_{m=1}^{K}{E}_{\substack{{{{v}_{n}}\sim {{P}_{n(v)}}}\\{{{{r}_{l}}\sim {{P}_{l(r)}}}}}}\sigma( ( \vec{u}{{_{i}^{'}}^{T}}\!\cdot\! {{{\vec{u}}}_{{{r}_{s}}}}\!\cdot\! {{{\vec{u}}}_{{{r}_{s}}}}^{T}\!\cdot\! {{{\vec{u}}}_{j}} )\!+\!( \vec{u}{{_{i}^{'}}^{T}}\!\cdot\! {{{\vec{u}}}_{{{r}_{l}}}}\!\cdot\! {{{\vec{u}}}_{{{r}_{l}}}}^{T}\!\cdot\!{{{\vec{u}}}_{n}} ))\Big)\\
\end{split}
\end{equation}
\begin{equation}
\label{updateurt}
\begin{split}
&\frac{\partial {O}}{\partial {{{\vec{u}}}_{r_t}}}={{\omega }_{ij}}*{{\omega }_{ik}}*(2\vec{u}{{_{k}^{'}}^{T}}\cdot {{{\vec{u}}}_{{{r}_{t}}}}\cdot {{{\vec{u}}}_{i}})\\
&*\Big(1-\sigma( ( \vec{u}{{_{j}^{'}}^{T}}\cdot {{{\vec{u}}}_{{{r}_{s}}}}\cdot {{{\vec{u}}}_{{{r}_{s}}}}^{T}\cdot {{{\vec{u}}}_{i}} )+( \vec{u}{{_{k}^{'}}^{T}}\cdot {{{\vec{u}}}_{{{r}_{t}}}}\cdot {{{\vec{u}}}_{{{r}_{t}}}}^{T}\cdot {{{\vec{u}}}_{i}} ))\Big)\\
&+{{\omega }_{ji}}*{{\omega }_{ik}}*(2\vec{u}{{_{k}^{'}}^{T}}\cdot {{{\vec{u}}}_{{{r}_{t}}}}\cdot {{{\vec{u}}}_{i}})\\
&*\Big(1-\sigma( ( \vec{u}{{_{i}^{'}}^{T}}\cdot {{{\vec{u}}}_{{{r}_{s}}}}\cdot {{{\vec{u}}}_{{{r}_{s}}}}^{T}\cdot {{{\vec{u}}}_{j}} )+( \vec{u}{{_{k}^{'}}^{T}}\cdot {{{\vec{u}}}_{{{r}_{t}}}}\cdot {{{\vec{u}}}_{{{r}_{t}}}}^{T}\cdot {{{\vec{u}}}_{i}} ))\Big)\\
&+{{\omega }_{ji}}*{{\omega }_{ki}}(2\vec{u}{{_{i}^{'}}^{T}}\cdot {{{\vec{u}}}_{{{r}_{t}}}}\cdot {{{\vec{u}}}_{k}})\\
&*\Big(1-\sigma( ( \vec{u}{{_{i}^{'}}^{T}}\cdot {{{\vec{u}}}_{{{r}_{s}}}}\cdot {{{\vec{u}}}_{{{r}_{s}}}}^{T}\cdot {{{\vec{u}}}_{j}} )+( \vec{u}{{_{i}^{'}}^{T}}\cdot {{{\vec{u}}}_{{{r}_{t}}}}\cdot {{{\vec{u}}}_{{{r}_{t}}}}^{T}\cdot {{{\vec{u}}}_{k}} ))\Big)
\end{split}
\end{equation}

The detailed optimization procedure is described in Algorithm 1. The embeddings for entities and relationships are all randomly initialized at first. Then, during each iteration, all nodes will be selected for optimization. In each phase, the vector embedding for current node and its neighbor nodes with relations will be updated using negative sampling method. The algorithm will be stopped until convergence.

 \begin{algorithm}
 \caption{MNE: Multi-relational Network Embedding}
 \label{alg1}
 \begin{algorithmic}[1]
 \REQUIRE  multi-relational network $G=( V,E )$,  $\eta $ : learning rate, K : $\#$ of negative samples, D : the dimensionality
 \ENSURE  representations of nodes and relations \\ $\Theta=\{{{\{ {{{\vec{u}}}_{i}} \}}_{i=1\ldots |V|}}$, ${{\{ \vec{u}_{i}^{'} \}}_{i=1\ldots |V|}}$, ${{\{ {{{\vec{u}}}_{{{r}_{i}}}} \}}_{i=1\ldots |R|}}\}$
 \STATE Randomly initialize $\Theta$
 \REPEAT
 \STATE Sample one node ${{v}_{i}}$ from $V$\\
 \STATE Sample ${v}_{i}$'s neighboring node ${{v}_{j}}$ with relation ${{r}_{s}}$, ${v}_{i}$'s neighboring node ${{v}_{k}}$ with relation ${{r}_{t}}$ 
 \STATE  Update $\Theta$ according to Eq.(15-22) for $MNE^+$ / Update $\Theta$ according to Eq.(24-31) for $MNE^*$
         \FOR{$m=0$ to $K$}
         \STATE Sample a negative node ${{v}_{n}}$ and a negative relation ${{r}_{l}}$\\
         \STATE JUpdate $\Theta$ according to Eq.(15-22) for $MNE^+$ / Update $\Theta$ according to Eq.(24-31) for $MNE^*$
         \ENDFOR
 \UNTIL{convergence}
 \STATE \textbf{return} ${\Theta}$
 \end{algorithmic}
 \end{algorithm}

\subsection{Time Complexity}
In this section, we show the time complexity of our proposed model is linear to the number of edges $| E |$ and independent on the number of nodes $| V |$. In practice, sampling a node or an edge takes constant time $O( 1 )$. Optimization with negative samples takes $O( d*(K+1) )$ time, where d is the dimension of the vector and K is the number of negative samples. For cases shown in Section \ref{sec:model_description}, the complexity is $O(3*d*(K+1))$. The number of steps need for the optimization is usually proportional to the number of edges $| E |$\cite{DBLP:conf/www/TangQWZYM15}. Therefore, the overall time complexity of our model is $O( d*K*|E| )$.
\begin{table}[h]
  \caption{Statistics of the datasets used for evaluation}
  \centering
  \label{table1}
  \begin{tabular}{ccccc}
    \toprule
		Dataset&\#Entity&\#Relation&\#Triplet& \#Tri-nodes\\
		\midrule
		\multirow{2}{*}{WN18}&\multirow{2}{*}{40943}&\multirow{2}{*}{18}&\multirow{2}{*}{151442}&895\\
		&&&&(2.19\%)\\
		\midrule
        \multirow{2}{*}{FB15K}&\multirow{2}{*}{14951}&\multirow{2}{*}{1345}&\multirow{2}{*}{592213}&6198\\
        &&&&(41.46\%)\\
	\bottomrule
\end{tabular}
\end{table}

\section{Experiment}
\begin{table*}[t]
	\newcommand{\tabincell}[2]{\begin{tabular}{@{}#1@{}}#2\end{tabular}}
    \renewcommand{\multirowsetup}{\centering}
	\centering
	\caption{Performance comparison on triplet classification}\label{tab:classification}
    \scalebox{1.2}{
	\begin{tabular*}{14.6cm}{c|c|c|c|c|c|c|c}
		\toprule
		\multirow{4}{*}{WN18}
        &Methods&${\!M\!N\!E}^+$&${\!M\!N\!E}^*$&LINE-1st-order&LINE-2nd-order&DeepWalk&RLine\\
		\cline{2-8}
		&\tabincell{c}{Acc.}&\tabincell{c}{\textbf{86.74\%}}& 78.02\% &\tabincell{c}{50.47\%}
        &\tabincell{c}{54.34\%}&\tabincell{c}{53.28\%}&\tabincell{c}{82.26\%}
        \\
        \cline{2-8}
        &Methods&TransE(bern)&TransE(unif)&TransH(bern)&TransH(unif)&TransR(bern)&TransR(unif)\\
        \cline{2-8}
        &\tabincell{c}{Acc.}&\tabincell{c}{81.31\%}&\tabincell{c}{80.42\%}&\tabincell{c}{81.44\%}
        &\tabincell{c}{80.83\%}&\tabincell{c}{80.43\%}&\tabincell{c}{80.73\%}\\
 		\toprule

        \multirow{4}{*}{FB15K}
        &Methods&${\!M\!N\!E}^+$&${\!M\!N\!E}^*$&LINE-1st-order&LINE-2nd-order&DeepWalk&RLine\\
		\cline{2-8}
        &\tabincell{c}{Acc.}&\tabincell{c}{\textbf{90.08\%}}& 75.95\% &\tabincell{c}{58.67\%}
        &\tabincell{c}{70.52\%}&\tabincell{c}{69.31\%}&\tabincell{c}{86.41\%}
        \\
		\cline{2-8}
        &Methods&TransE(bern)&TransE(unif)&TransH(bern)&TransH(unif)&TransR(bern)&TransR(unif)\\
        \cline{2-8}	
        &\tabincell{c}{Acc.}&\tabincell{c}{70.46\%}&\tabincell{c}{71.40\%}&\tabincell{c}{71.72\%}
        &\tabincell{c}{70.98\%}&\tabincell{c}{70.49\%}&\tabincell{c}{71.48\%}\\
        \bottomrule

		
	\end{tabular*}}
\end{table*}

\begin{table*}[t]
	\newcommand{\tabincell}[2]{\begin{tabular}{@{}#1@{}}#2\end{tabular}}
    \renewcommand{\multirowsetup}{\centering}
	\centering
	\caption{Performance comparison on link prediction}\label{tab:link_prediction}
    \scalebox{1.2}{
	\begin{tabular*}{14.6cm}{c|c|c|c|c|c|c|c}
		\toprule
		\multirow{4}{*}{WN18}
		&Methods&${\!M\!N\!E}^+$&${\!M\!N\!E}^*$&LINE-1st-order&LINE-2nd-order&DeepWalk&RLine\\
		\cline{2-8}
		&\tabincell{c}{Acc.}&\tabincell{c}{\textbf{85.04\%}}& 76.51\% &\tabincell{c}{50.94\%}
        &\tabincell{c}{54.12\%}&\tabincell{c}{54.54\%}&\tabincell{c}{83.42\%}\\
        \cline{2-8}
        &Methods&TransE(bern)&TransE(unif)&TransH(bern)&TransH(unif)&TransR(bern)&TransR(unif)\\
        \cline{2-8}
        &\tabincell{c}{Acc.}&\tabincell{c}{82.76\%}&\tabincell{c}{82.46\%}&\tabincell{c}{83.48\%}
        &\tabincell{c}{82.22\%}&\tabincell{c}{82.36\%}&\tabincell{c}{82.38\%}\\
        \toprule
        \multirow{4}{*}{FB15K}
        &Methods&${\!M\!N\!E}^+$&${\!M\!N\!E}^*$&LINE-1st-order&LINE-2nd-order&DeepWalk&RLine\\
		\cline{2-8}
        &\tabincell{c}{Acc.}&\tabincell{c}{\textbf{91.81\%}}& 75.95\% &\tabincell{c}{59.27\%}
        &\tabincell{c}{64.13\%}&\tabincell{c}{69.55\%}&\tabincell{c}{86.86\%}\\
		\cline{2-8}
        &Methods&TransE(bern)&TransE(unif)&TransH(bern)&TransH(unif)&TransR(bern)&TransR(unif)\\
        \cline{2-8}
        &\tabincell{c}{Acc.}&\tabincell{c}{69.40\%}&\tabincell{c}{71.23\%}&\tabincell{c}{69.77\%}
        &\tabincell{c}{72.46\%}&\tabincell{c}{71.35\%}&\tabincell{c}{71.77\%}\\
        \bottomrule
		
	\end{tabular*}}
\end{table*}
To evaluate the performance of the proposed multi-relational network embedding (MNE), we employ two well-known benchmark datasets, namely, WN18 and FB15K which are extracted from the real-world multi-relational networks WordNet \cite{DBLP:journals/cacm/Miller95} and Freebase \cite{DBLP:conf/sigmod/BollackerEPST08} respectively. Table \ref{table1} tabulates their statistics where the tri-nodes refers to the nodes conforming a triangular structure in networks. We compare our proposed MNE with several existing methods in trans-family, including TransE, TransH and TransR
where the two settings ``unif'' and ``bern'' to sample negative instances are used for the embedding learning \cite{DBLP:conf/aaai/LinLSLZ15}.

We also compare our proposed approach with the state-of-the-art approaches for network embedding, including DeepWalk 
and LINE. \footnote{As LINE and Deepwalk can only deal with single relational networks, we treat the linkages of various types between two nodes in multi-relational networks as a weighted single relation.}
LINE and Deepwalk, the representation algorithms for single relational networks, rely on the weight of the edge between nodes during the learning process. To adapt LINE and Deepwalk to multi-relational networks, in our experiments, we utilize the number of categories of relations between two nodes as the weight of the edges. In our experiments, both first-order proximity and second-order proximity terms in LINE are investigated for comparison, denoted as LINE-1st-order and LINE-2nd-order respectively.

Furthermore, we extend the conventional LINE by incorporating the representations of different labels of relations. For example, we revise the LINE-2nd-order by taking Eq.(\ref{eq:p_line}) in place of the probability of ``context'' $v_j$ generated by $v_i$ and we call the revised model as RLine in the sequel.
\begin{eqnarray}
\label{eq:p_line}
{{p}}( v_j, r_s | v_i )=\frac{\exp ( \vec{u}{{_{j}^{'}}^{T}}( {{{\vec{u}}}_{i}}+{{{\vec{u}}}_{{{r}_{s}}}} ) )}{\sum\limits_{( {{v}_{i}},{{r}_{p}},{{v}_{x}} )\in E^{'}}{\exp ( \vec{u}{{_{x}^{'}}^{T}}( {{{\vec{u}}}_{i}}+{{{\vec{u}}}_{{{r}_{p}}}} ) )}} \label{multi}
\end{eqnarray}

\begin{figure*}
	\centering
	\subfigure[Triplet Classificati on WN18]{\label{dim1}\includegraphics[width=0.2\paperwidth]{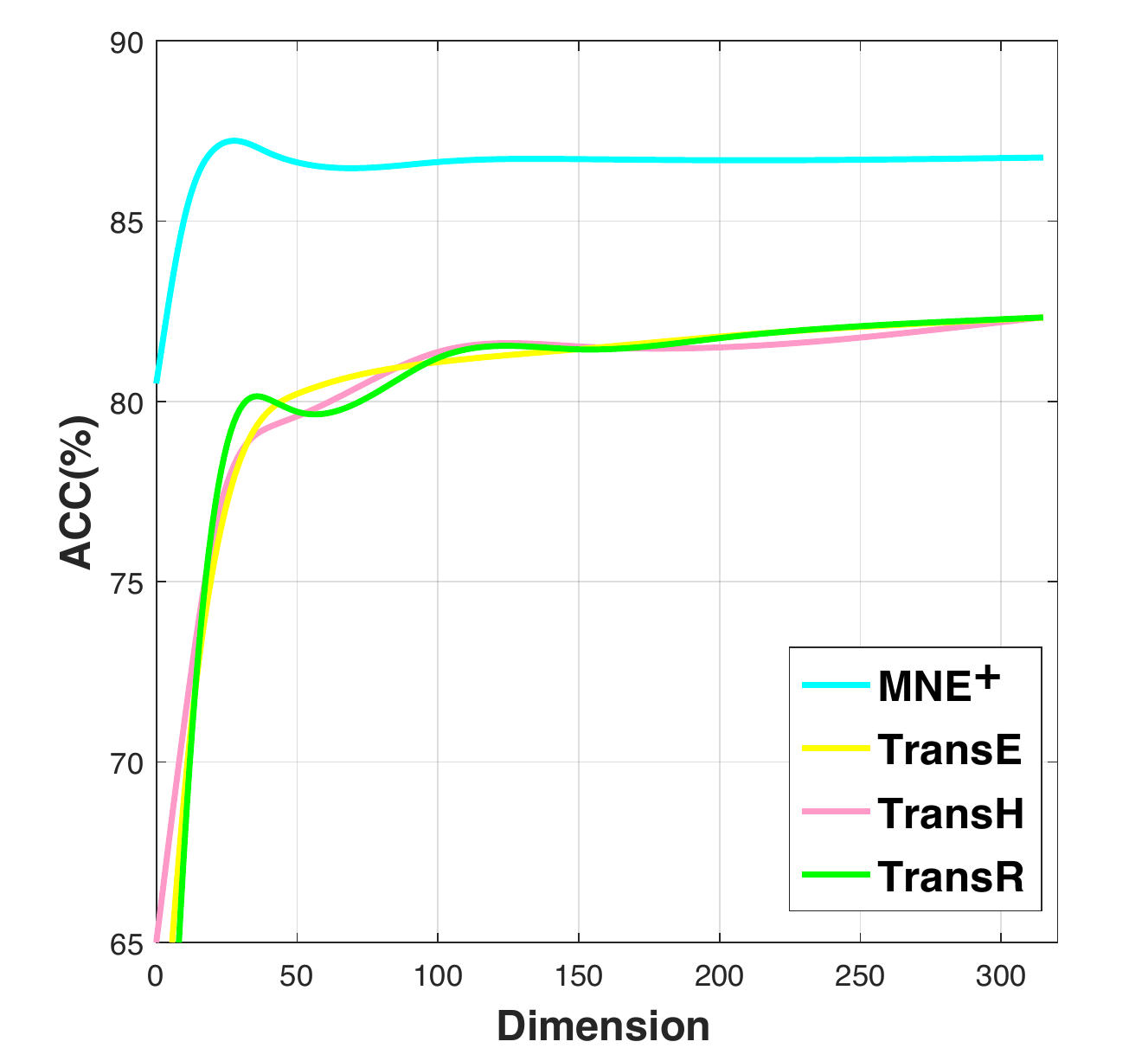}}
	\subfigure[Triplet Classificati on FB15K] {\label{dim2}\includegraphics[width=0.2\paperwidth]{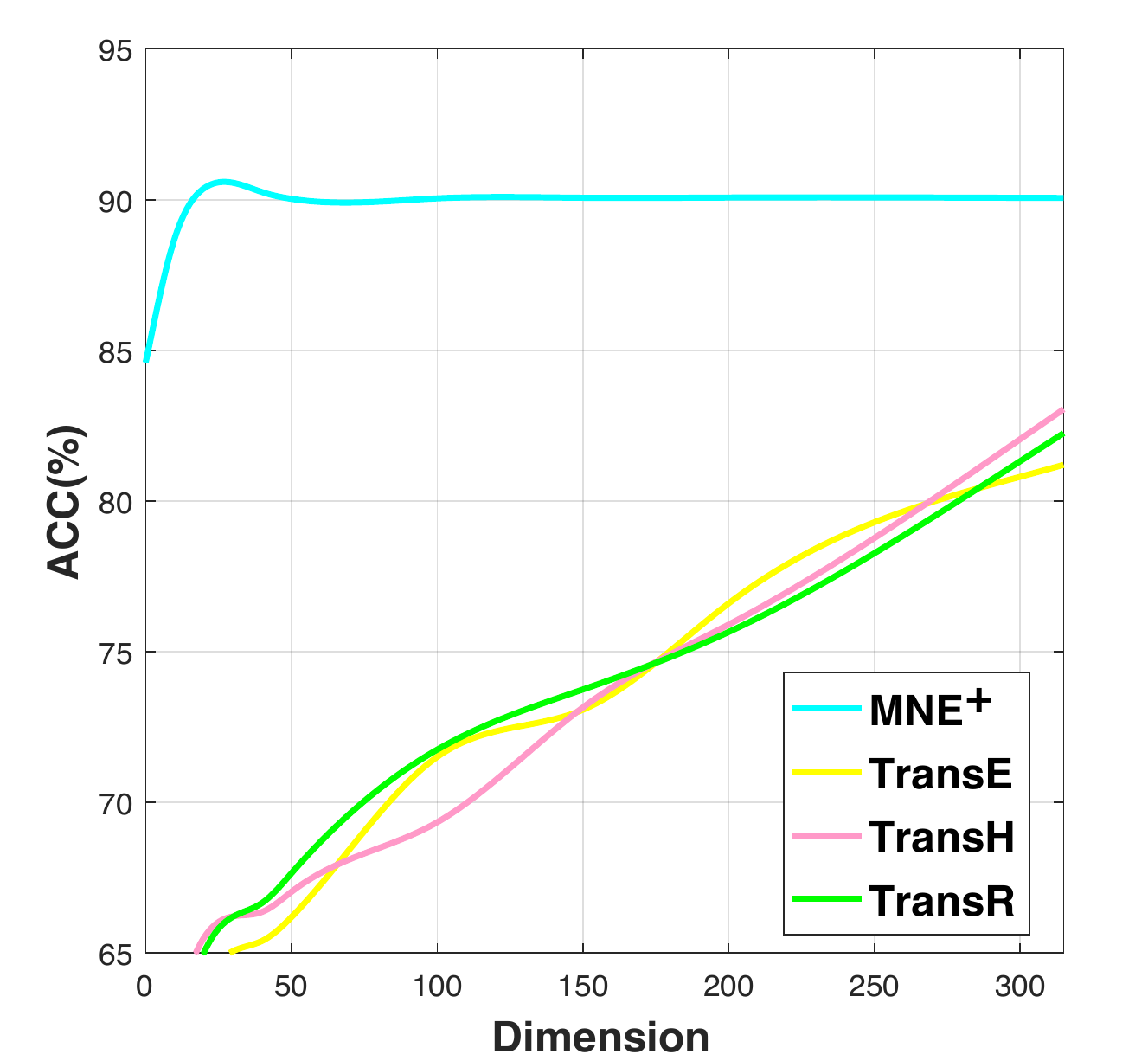}}
	\subfigure[Link Prediction on WN18] {\label{dim3}\includegraphics[width=0.2\paperwidth]{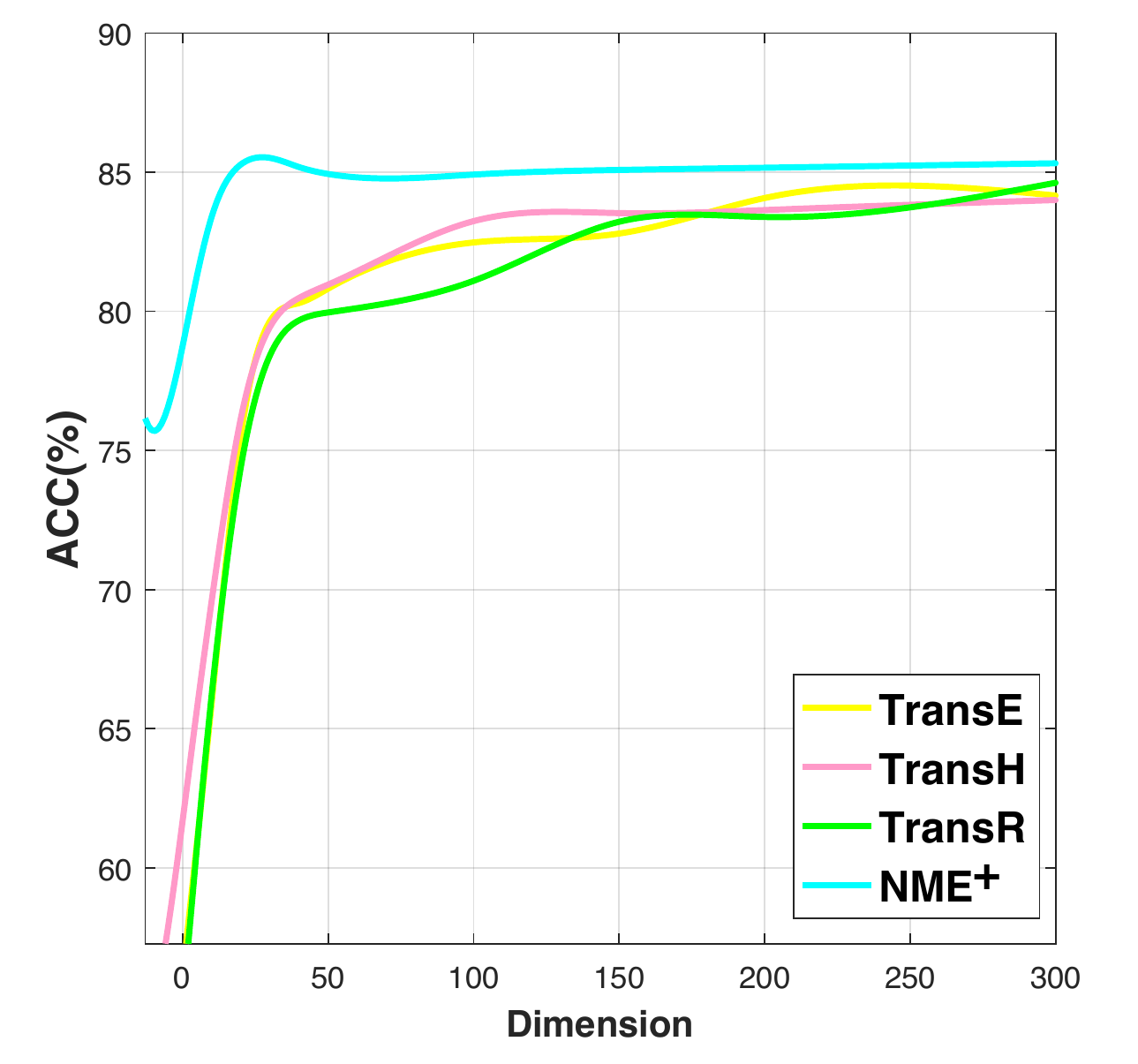}}
	\subfigure[Link Prediction on FB15K] {\label{dim4}\includegraphics[width=0.2\paperwidth]{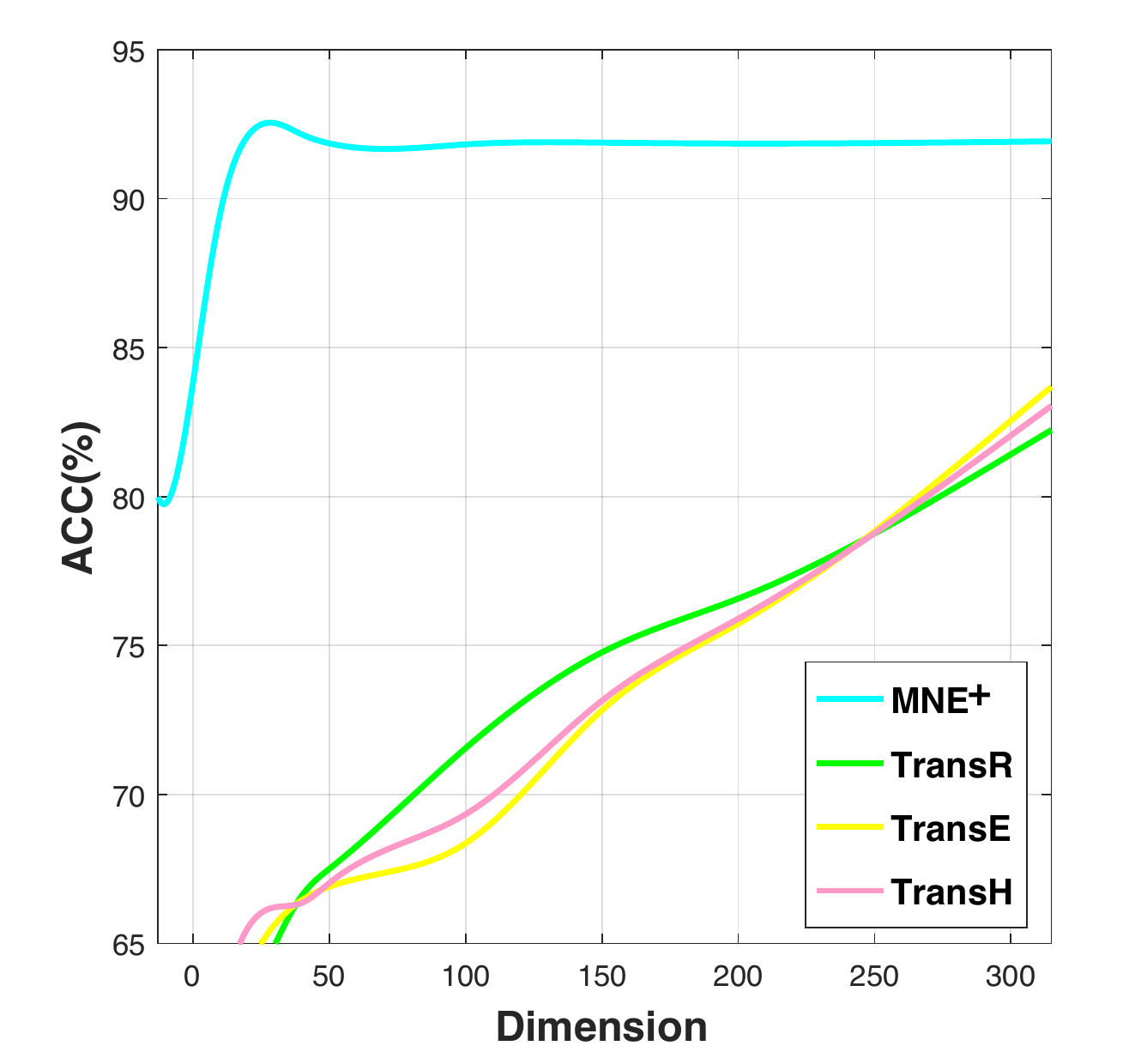}}
	\caption{ACC v.s. dimension}
    \label{fig:dimension}
\end{figure*}

\subsection{Triplet Classification}


The triplet classification task has been widely investigated for the performance evaluation of representation learning approaches, which is usually translated into a binary classification task to judge whether a given triplet is a fact or not in a given knowledge base.\\
\textbf{Evaluation Protocol}
In this task, we perform binary classification as in \cite{DBLP:conf/kdd/GroverL16}. The embeddings of networks are first obtained via our proposed models and the comparison models on each entire dataset, then evaluated by the binary classifier. The triplet facts $\left( h,r,t \right)$ appeared in the dataset are taken as the positive samples. And we randomly sampled the same number of triplets that have not appeared in the dataset as the negative triplets. We concatenate the obtained low-dimensional vectors of the head entity, relation and tail entity as the input of a classifier. The training set and test set are randomly splited in a ratio of $x\%:1-x\%$. We use the classification accuracy as the evaluation criterion. And both logistic regression (LR) and support vector machine (SVM) are adopted for the classifier with similar results achieved. We adopt LR for its efficiency in this paper. 

\textbf{Results} Table \ref{tab:classification} shows the performance comparison among the existing approaches for triplet classification. We observe that:

\begin{inparaenum}[(1)]
\item The proposed MNEs (${\!M\!N\!E}^+$, ${\!M\!N\!E}^*$) and the trans-family perform consistently better than the network embedding methods (i.e. DeepWalk and LINE) which treat the relations semantically indistinguishable; 

\item For both benchmark datasets, our proposed approach ${\!M\!N\!E}^+$ outperforms all the baseline methods, but ${\!M\!N\!E}^*$ failed in WN18. The reason behind is that the bridging function by adopting the product operation does not distinguish between source vector $\vec{u}_i$ and target vector $\vec{u}_i'$, owing to the product of $\vec{u}_{r_s}\cdot\vec{u}_{r_s}^T$ being a symmetric matrix. Let $H = \vec{u}_{r_s}\cdot\vec{u}_{r_s}^T$, we have $\vec{u}_j'^T\cdot H\cdot\vec{u}_i=\vec{u}_i^T\cdot H\cdot\vec{u}_j'$. The left side of the equation indicates that $\vec{u}_i$ and $\vec{u}_j'$ are the representation of $v_i$ being the source node and $v_j$ being the target node respectively (Recall from Fig.\ref{fig:directedEdge}). While the right side of the equation indicates that $\vec{u}_j'$ and $\vec{u}_i$ are representing $v_j$ as the source node and $v_i$ as the target node respectively, which in fact is against our original intention of using two sets of embeddings $u$ and $u'$ to distinguish the roles of a node. Thus, we conclude that the bridging function of the product operation may compromise the performance of ${\!M\!N\!E}^*$ on directed networks. For ${\!M\!N\!E}^+$ which adopts the addition bridging function, $u_i$ and $u_i'$ can play well different roles (being source or tail) in a directed edge, as $\vec{u}_j'^T(\vec{u}_i + \vec{u}_{r_s}) \neq \vec{u}_i^T(\vec{u}_j' + \vec{u}_{r_s})$, which explains ${\!M\!N\!E}^+$'s being superior to ${\!M\!N\!E}^*$ ; 
 

\item The trans-family does not work well on FB15K while our proposed MNEs can still achieve high accuracy. And ${\!M\!N\!E}^*$ also performs better than trans-family. As reported in Table \ref{table1}, FB15K is a far more dense multi-relational network with more relation types than WN18. The relation-specific local structures are intuitively more complex. And in FB15K dataset, there are more nodes with the triangular structures compared to WN18. That accounts for the performance degradation of trans-family enforcing the constraints of $h+r=t$. 

\item The performance achieved by RLine on two datasets has been greatly improved compared with LINE. It is shown that introducing the edge labels into the networks plays a positive role in improving the performance of the representation learning algorithm. RLine performs better than ${M\!N\!E}^*$ but worse than ${M\!N\!E}^+$ which further validate the effectiveness of the addition bridging function and the importance of capturing the parallelogram structures. \footnote{Note that for the different proportion of train-test splits, the observations over all compared models are roughly same and consistent. Due to the page limit, only the experimental results in the dataset of 8:2 train-test split are detailed in Table \ref{tab:classification}.}

\end{inparaenum}



\subsection{Link Prediction}

Link prediction is to predict the missing h or t for a triplet fact $\left( h,r,t \right)$ in a given KG. That is to obtain the best answer of t given $\left( h,r \right)$ or to obtain the best answer of h given $\left( r,t \right)$. \\
\textbf{Evaluation Protocol} Again, the link prediction problem can be posed as a binary classification problem by employing the low-dimensional vectors obtained from our proposed model. While the triplets in a KG can form the positive samples, the negative samples can be generated by corrupting each triplet of fact $\left( h,r,t \right)$ with the head (h) or tail (t) replaced. The experiments are evaluated using 80/20 rule for the train-test split. During the embedding training process, only the training set is used. Note that the training dataset is forced to cover all nodes. Again, a LR classifier is trained by using the obtained low-dimensional vectors and tested on the corrupted edges. Compared to triplet classification, the training set for link prediction classifier is the same as embedding training set, and the test set will no longer included in the dataset for representation learning. Again, we use the classification accuracy as the evaluation criterion.
 

\textbf{Results}
The evaluation results are shown in Table \ref{tab:link_prediction}. We made similar observations as those for triplet classification. 
In particular, the proposed MNEs and the trans-family are performing obviously better than the network embedding methods on WN18. The trans-family methods do not perform well on FB15K. The phenomenon further confirms that the triangular structures in multi-relational networks will degrade the performance of the trans-family. Instead, MNEs perform better in FB15K than WN18, which further verifies the advantage of MNEs dealing with the networks with high triangular structure ratio. RLine performs consistently better than the LINE-1st-order and LINE-2nd-order on two datasets, indicating the importance of the label of edges for network representation learning and the effectiveness of adopting the bridging function of addition. ${M\!N\!E}^+$ outperforms all the other methods on both WN18 and FB15K consistently.


\subsection{Model Sensitivity}
Among the methods proposed for multi-relational networks, we also compare their performances on the triplet classification and link prediction (WN18 and FB15K) under the settings of different dimensions of the representation. Here we refer ${M\!N\!E}^+$ as the representative of our proposed model MNE. The results are shown in Fig.\ref{fig:dimension}. We observe that: 1) There is a positive correlation between the classification accuracy and the dimension. After reaching a specific dimension, the classification accuracy converges; 2) MNE outperforms other state-of-the-art methods for all the dimensionality settings. In particular, MNE can work very well even at a very low dimension (2 to 5); 3) MNE converges when the dimension reaches 20, while the other methods reach the good performance when the dimension is at least 100. We conclude that MNE could obtain a more compact representation compared with other approaches. Besides, similar to LINE, we adopt the negative sampling to substantially reduce the computational cost of learning, which allows MNE to scale up to the network of large size.
\begin{figure}[htb]
\centering\includegraphics[width=0.4\paperwidth]{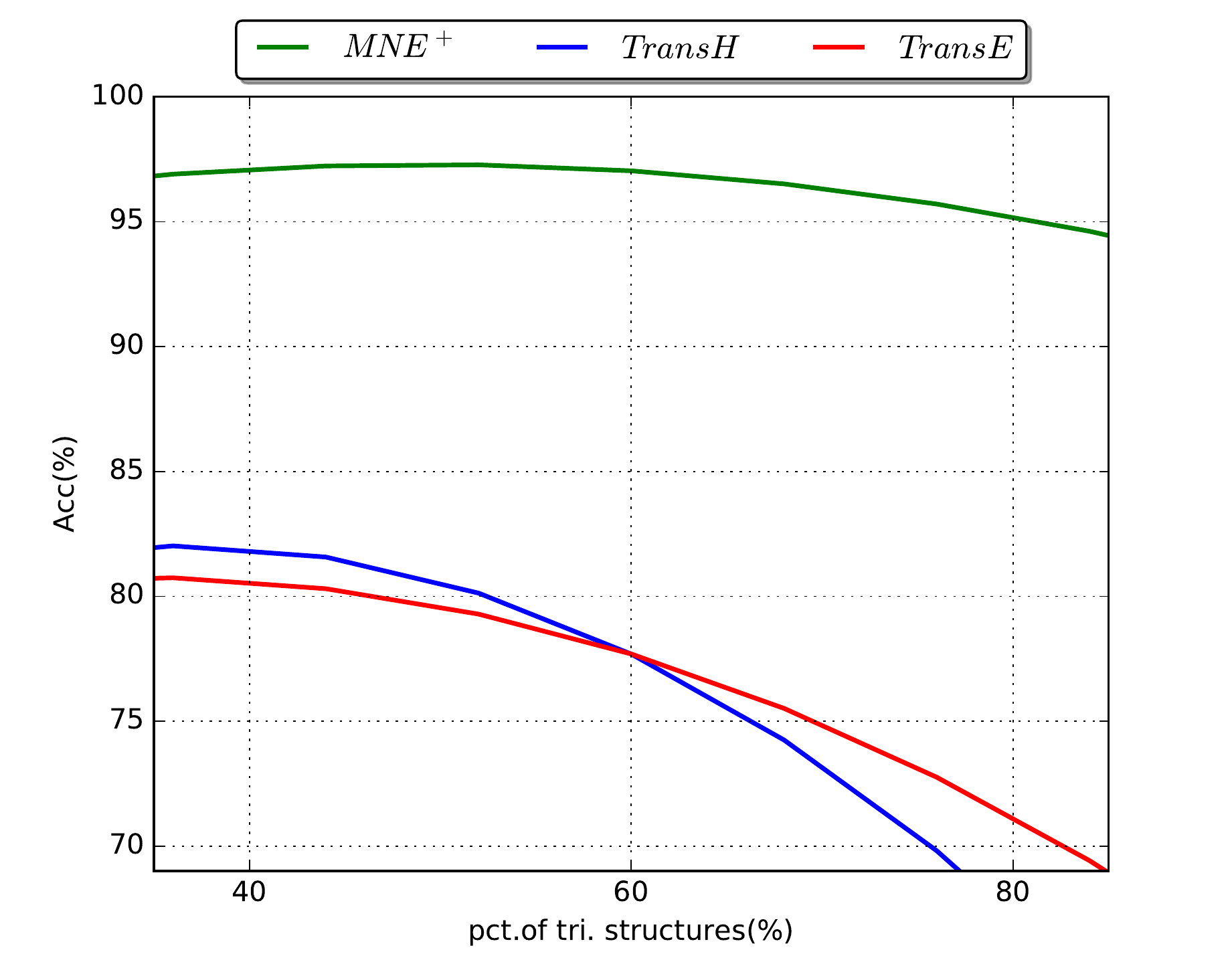}
\caption{Parallelogram structure examples.}
\label{fig:tri_rho}
\end{figure}

To further evaluate whether the proposed MNE alleviates the limitation of triangular connectivity structures, we conduct triplet classification experiments on WN18 dataset with different triangular proportions. The nodes and edges which do not belong to a triangular structure are gradually added to simulate the deceasing number of the triangular structures. Fig.\ref{fig:tri_rho} pans out as we expected, the accuracy of the link prediction obtained by TransE and TransH decreases as the number of the triangular structures increases. And our model MNE is relatively stable at a high level of accuracy when the percentage of tri-nodes goes up.





\section{Conclusion}
In this paper, we propose a novel multi-relational network embedding model. Many existing knowledge graph embedding methods share an intrinsic limitation of adopting a hard constraint on the inferred embedding. By defining an objective function which can implicitly preserve triangular and parallelogram structures, the proposed model can give more flexible embedding results. Negative sampling are used to reduce the computational cost for the learning process. The extensive experiments conducted on two real world datasets demonstrate that our proposed model outperforms a number of state-of-the-art embedding methods. This paper only explores the local structures to obtain embedding without considering other information carried in the network. We would like to explore the idea of incorporating semantic information in our framework for the future work.

\bibliographystyle{IEEEtran}
\bibliography{extendaaai}

\end{document}